\documentclass{article}
\usepackage[utf8]{inputenc}
\usepackage[pdftex]{color,graphicx}
\usepackage{graphicx}
\usepackage{amsfonts}
\usepackage{amssymb}
\usepackage{url}
\usepackage{bm}
\usepackage[compress]{cite}
\usepackage{makecell}

\usepackage[dvipsnames]{xcolor}
\definecolor{darkgreen}{rgb}{0,0.65,0}
\usepackage{slashed}
\usepackage[linktoc=all,colorlinks=true,linkcolor=black,urlcolor=magenta,citecolor=black]{hyperref}
\hypersetup{%
  colorlinks = true,
  linkcolor  = black
}
\definecolor{rossoCP3}{cmyk}{0,.88,.77,.40}
\definecolor{verdeCP3}{rgb}{0.09765625, 0.57421875, 0.1015625}
\definecolor{bluCP3}{rgb}{0, 0.23, 0.67}
\ifx\pdfoutput\undefined
\usepackage[dvips,bookmarks]{hyperref}    % This is for arXiv.org
\else
\usepackage{hyperref}    % This is for pdftex
\fi
\hypersetup{colorlinks, bookmarksopen, bookmarksnumbered,
citecolor=verdeCP3, linkcolor=bluCP3, pdfstartview=FitH, urlcolor=rossoCP3}

\newcommand{\Fig}[1]{Fig.~\ref{#1}}
\newcommand{\Tab}[1]{Tab.~\ref{#1}}
\newcommand{\Sec}[1]{Sec.~\ref{#1}}

\newcommand{\be}{\begin{equation}}
\newcommand{\ee}{\end{equation}}
\newcommand{\de}{\mathrm d}
\newcommand{\nn}{\mathcal{N}}
\def\bea{\begin{eqnarray}}
\def\eea{\end{eqnarray}}
\definecolor{corr}{rgb}{1,0,.2}

\title{Searching for Sterile Neutrino with X-ray Intensity Mapping}
\author{regis.mrc }
\date{September 2019}

\begin{document}
\begin{center}
{\LARGE\color{black}\bf Searching for Sterile Neutrino with X-ray Intensity Mapping\\[1mm] }

\medskip
\bigskip\color{black}\vspace{0.6cm}

{
{\large\bf Andrea Caputo$^{a}$,}\ 
{\large\bf Marco Regis$^{b,c}$,}\ 
{\large\bf Marco Taoso$^{c}$,}\ 
}
\\[7mm]

{\it $^a$  Instituto de F\'{i}sica Corpuscular, CSIC-Universitat de Valencia,
Apartado de Correos 22085, E-46071, Spain
}\\[3mm]
{\it $^b$  Dipartimento di Fisica, Universit\`{a} di Torino, via P. Giuria 1, I--10125 Torino, Italy}\\[3mm]
{\it $^c$  Istituto Nazionale di Fisica Nucleare, Sezione di Torino, via P. Giuria 1, I--10125 Torino, Italy}\\[3mm]
\end{center}

\bigskip

\centerline{\large\bf Abstract}
\begin{quote}
\color{black}\large 
The cosmological X-ray emission associated to the possible radiative decay of sterile neutrinos is composed by a collection of lines at different energies. For a given mass, each line corresponds to a given redshift.
In this work, we cross correlate such line emission with catalogs of galaxies tracing the dark matter distribution at different redshifts. We derive observational prospects by correlating the X-ray sky that will be probed by the eROSITA and Athena missions with current and near future photometric and spectroscopic galaxy surveys. A relevant and unexplored fraction of the parameter space of sterile neutrinos can be probed by this technique.

\end{quote}

\newpage

\section{Introduction}
\label{sec:introduction}
Unveiling the nature of dark matter (DM) is one of the most intriguing goals of fundamental physics nowadays. In fact, despite the compelling astrophysical and cosmological evidences, still we do not know what DM is made of.
The most popular DM candidates are those connected to additional problems faced by the Standard Model (SM) of particle physics.

An example is the sterile neutrino. This particle is a singlet under the SM gauge group and arises in scenarios of active neutrino mass generation, notably the see-saw mechanism~\cite{Minkowski:1977sc,Mohapatra:1979ia}.
The mass of the sterile neutrino can span a very large range, from the GUT scale \cite{Fukugita:1986hr,Brahmachari:1998kt,Fritzsch:1974nn,Nandi:1985uh,Mohapatra:1986bd} down to eV \cite{Diaz:2019fwt,Dentler:2018sju,Kopp:2011qd}. In particular, a sterile neutrino with a mass around the keV scale has attracted a lot of interest since it constitutes a viable and appealing DM candidate (see, e.g., the  $\nu$MSM model \cite{Asaka:2005an} for a concrete implementation and Refs.~\cite{Adhikari:2016bei,Abazajian:2017tcc,Boyarsky:2018tvu} for reviews).

A keV sterile neutrino mixes with ordinary active neutrinos and can be produced in the early Universe through oscillations (Dodelson-Widrow mechanism)~\cite{Dodelson:1993je}, resonantly enhanced oscillations in presence of a primordial lepton asymmetry (Shi-Fuller mechanism)~\cite{Shi:1998km}, through 
the decay of heavy particles~\cite{Kusenko:2006rh,Petraki:2007gq,Khalil:2008kp,Merle:2013wta,Shuve:2014doa,Abada:2014zra,Konig:2016dzg, Merle:2015oja,Caputo:2018zky,Kaneta:2016vkq} or obtain the correct DM abundance via
dilution of a thermal sterile neutrino component through entropy production~\cite{Asaka:2006ek,Bezrukov:2009th,Nemevsek:2012cd,Patwardhan:2015kga}~\footnote{See~\cite{Gelmini:2019clw,Gelmini:2019wfp} for the production of sterile neutrinos in non-standard cosmologies.}.

As a basic requirement in order to be the DM, sterile neutrinos need to be cold enough to be confined inside galaxies.
Fermions obey the Pauli exclusion principle and in a system with $N$ fermions, the minimum momentum is therefore $p \sim N^{1/3}h/R$, where $h$ is the Planck constant and $R$ the size of the fermionic system.
Analyzing the DM phase-space distribution of galaxies one can deduce a lower limit on the mass of a fermionic DM candidate, the so-called Tremaine–Gunn-type bound \cite{Tremaine:1979we}.
Additional and stronger constraints on the coldness of DM can be obtained from the number of collapsed structures (like counts of the Milky Way satellites) and measurements of the matter power spectrum at small scales, in particular through the Lyman-alpha forest method.
The corresponding bounds on the sterile neutrino mass depend on the production mechanism. We refer to Refs.~\cite{Adhikari:2016bei,Abazajian:2017tcc,Boyarsky:2018tvu} for more details and discussions on the uncertainties affecting these astrophysical constraints.

Sterile neutrinos decay into SM particles via the mixing with the active neutrinos, the main process being $\nu_{s}\rightarrow \nu\nu\nu $.
An additional channel is the radiative decay $\nu_{s}\rightarrow \nu\gamma$, occurring with a rate \cite{Lee:1977tib, Pal:1981rm}: 
\begin{equation}
   \Gamma_S \equiv \Gamma_{\nu_s\rightarrow \nu \gamma} \sim (7.2\,\, 10^{29}{\rm s})^{-1}\Big(\frac{\sin^2(2\theta)}{10^{-8}}\Big) \Big(\frac{m_S}{1\,{\rm keV}}\Big)^5\;.
\end{equation}
where $\theta$ is the mixing angle between active and sterile neutrinos, and $m_S$ is the mass of the latter. 

Therefore, a smoking gun signature for sterile neutrino DM would be to detect a monoenergetic X-ray signal from the above process. 
Constraints on the sterile-active neutrino mixing have been set from the non-observation
of such decay line from different targets, including dwarf spheroidal galaxies, clusters of galaxies, the Milky Way and the X-ray background (again see, e.g., reviews in Refs.~\cite{Adhikari:2016bei,Abazajian:2017tcc,Boyarsky:2018tvu} and references therein). 
Interestingly enough, the detection of an unidentified line at energy $E \simeq 3.5$ keV was reported recently in different astrophysical environments, including galaxy clusters, with both stacked \cite{Bulbul:2014sua} and individual \cite{Bulbul:2014sua,Boyarsky:2014jta} spectra, the Andromeda galaxy \cite{Boyarsky:2014jta} and the Galactic Center \cite{Boyarsky:2014ska,Riemer-Sorensen:2014yda}. It has been suggested that this observational finding may be a signature of the decay of a sterile-neutrino DM. Several works have been trying to test the presence of such excess, finding controversial results (see, e.g., Ref.~\cite{Dessert:2018qih} and the review in Ref.~\cite{Adhikari:2016bei}). Future data and new analyses are therefore necessary to reach a conclusive answer.

In this work we entertain the possibility to detect the sterile neutrino decay signal produced in cosmic structures. This cumulative emission is the superposition of all the X-ray lines produced in DM halos and redshifted by the expansion of the Universe.
In order to exploit the correlation between the energy of the line and the redshift of the corresponding halos, we cross correlate the X-ray extragalactic emission with catalogs of galaxies, tracing the DM distribution at different redshifts.   
We derive prospects for the eROSITA mission~\cite{Merloni:2012uf}, currently in operation, expanding over the study conducted in Ref.~\cite{Zandanel:2015xca}. A significant improvement in the sensitivity will be provided by the next-generation X-ray telescope Athena~\cite{Nandra:2013shg}. We compute forecasts for its operations, and considering, on the galaxy catalog side, current and near future spectroscopic and photometric surveys.
In particular, Athena will allow to perform high-resolution spectroscopy thanks to the X-IFU instrument. The combination with spectroscopic galaxy catalogs will provide a framework where to fully perform a line intensity mapping analysis. Indeed, the Athena X-IFU detector will be able to resolve the narrow line induced by sterile neutrino decay and will provide a good rejection of backgrounds, given by continuum spectra and other emission lines at different energies, from the vast number of cross energy-redshift bins where the DM signal is absent.

The role of line intensity mapping for DM searches has been highlighted also in Ref.~\cite{Creque-Sarbinowski:2018ebl}.

The paper is organized as follows. \Sec{sec:models} describes the formalism used to compute the cross-correlation signal. In~\Sec{sec:exp} we describe the experimental configurations considered in the derivation of the projected bounds. The statistical analysis and results are presented in~\Sec{sec:res}. Details on the astrophysical models are in the Appendix. In \Sec{sec:conclusions} we conclude.

\section{Models}

\label{sec:models}

The monopole of the intensity associated to X-ray emission $I_\mathrm{X}$ can be described by means of an integral of the window function $W_\mathrm{X}$ as
\be
      I_\mathrm{X}(E) =
     \int_0^\infty d\chi
     ~W_\mathrm{X}(E, z)\ ,
  \label{eq:intensity}
\ee
where $E$ is the observed X-ray energy, and $\chi(z)$ is the comoving distance to redshift $z$, obeying, in a flat Universe, $c\, \de z/\de\chi=H(z)$ with $H(z)$ being the Hubble rate.
It is clear that $W_\mathrm{X}$ (also called weight function) provides the fraction of intensity emitted in a given redshift slice.

We will consider four extragalactic emitters of X-rays: Active Galactic Nuclei (AGN), galaxies, clusters of galaxies, and DM in the form of sterile neutrinos.

In the case of X-ray emission coming from the decay of sterile neutrinos DM, the window function takes the form
\be
 W_{X_S}(E, z) = \frac{\Omega_{\rm DM} \rho_c}{4\pi}\frac{\Gamma_S}{m_S(1+z)}
  \frac{1}{\sqrt{2\pi}\sigma_{E}}\exp\left[-\frac{(E - \frac{m_S}{2(1+z)})^2}{2\sigma_{E}^2}\right]\;,
  \label{eq:Wnu}
\ee
where the Gaussian function provides the broadening of the emission line (centered at $E=\frac{m_S}{2(1+z)}$) due to the spectral resolution $\sigma_{E}$ of the X-ray telescope.
We neglect the velocity dispersion of sterile neutrinos in halos, since it is always smaller than the experimental energy resolution considered (as we will comment in the following).

Throughout the paper, we consider the sterile neutrino to be the only DM component in the Universe, thus having a relic density set by $\Omega_{\rm DM} \rho_c$. We take the value of cosmological parameters from Ref.~\cite{Aghanim:2018eyx}.

\begin{figure}[t]
\begin{center}
\includegraphics[width= 0.48 \textwidth]{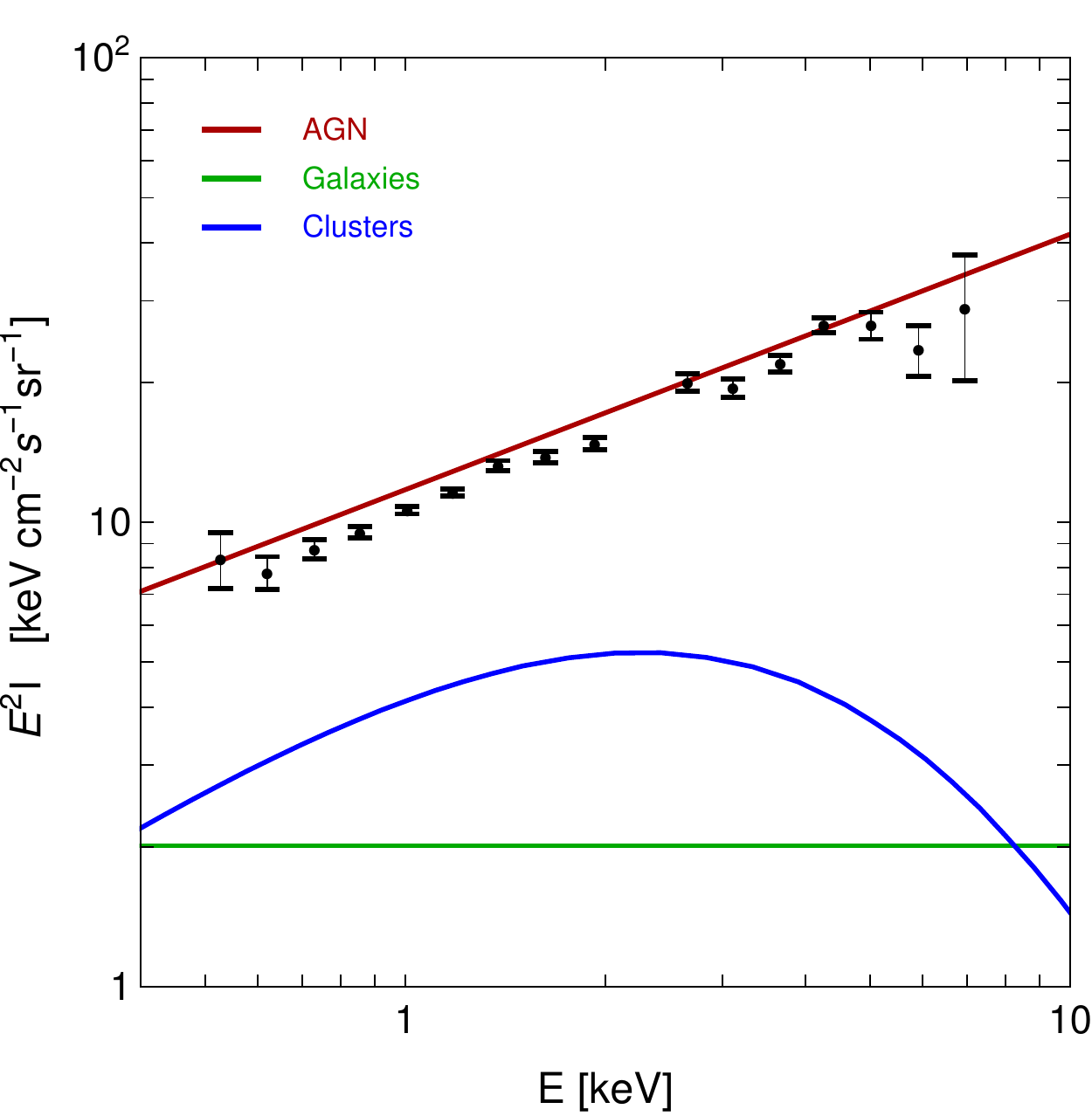}
\caption{Total X-ray intensity produced by AGN, galaxies and clusters (i.e., the sum of resolved and unresolved sources). Data points show the total extragalactic X-ray background  measured by Chandra~\cite{Cappelluti:2017miu}.}
\label{fig:intensity}
\end{center}
\end{figure}

For AGN and galaxies, $W_\mathrm{X}$ reads:
\be
W_\mathrm{X_a}(E, z)=\chi^2(z)\int^{L_{\rm max}(F_{sens},z)}_{L_{\rm min}} \de L\, \frac{dF}{dE}(L,z)\,\phi(L,z)\;,
\label{eq:Wastro}
\ee
where $F$ is the flux provided by a source with rest-frame luminosity $L$ at redshift $z$, $\phi$ is the X-ray luminosity function, namely, the number of X-ray sources per unit volume and unit luminosity, and the maximum luminosity $L_{\rm max}$ of an unresolved source is dictated by the sensitivity flux $F_{sens}$, providing the minimum detectable flux.
Details concerning the models adopted for $\phi$ are given in the Appendix.

Clusters of galaxies can emit X-rays by means of bremsstrahlung radiation of their gas. We assume the minimum mass of a cluster to be $M_{500}=10^{14}\,{\rm M_\odot}$, which is also approximately the same value one obtains from the temperature-mass relation \cite{Lieu:2015pit} by requiring to have a sizable emission above 1 keV (i.e., in the energy range we are considering).
We checked that all clusters above this mass will be under the detection reach of eROSITA and Athena, and are therefore masked in our analysis, see details in Appendix.

Note that by masking all halos above $10^{14}\,{\rm M_\odot}$, we will be masking as well the possible contribution of sterile neutrino decay from such massive halos.
In our computation, we thus set the maximum halo mass to be $M_{500}=10^{14}\,{\rm M_\odot}$ in the computation of the DM signal.
In order to retain instead such DM contribution, one could include clusters and treat them as a background component, as done in Ref.~\cite{Zandanel:2015xca}. This approach would increase the DM signal but at the cost of highly increasing the noise of the measurement.

The intensity provided by the astrophysical emitters is reported in \Fig{fig:intensity}, where we compare the predictions of the adopted models with the measurement of Ref.~\cite{Cappelluti:2017miu}.

\begin{figure}[t]
\begin{center}
\includegraphics[width= 0.47 \textwidth]{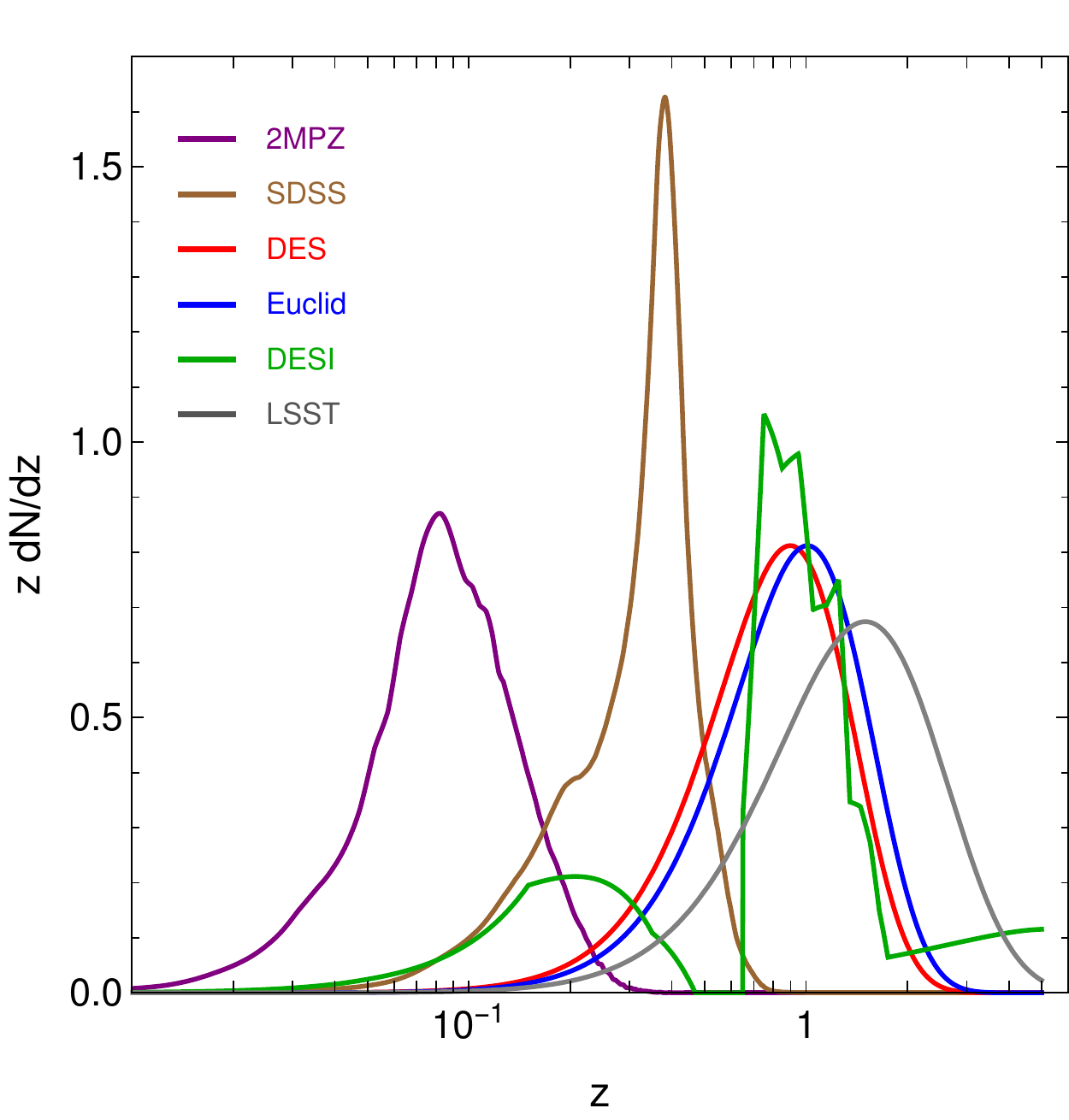}
\hspace{0.27cm}
\includegraphics[width= 0.49 \textwidth]{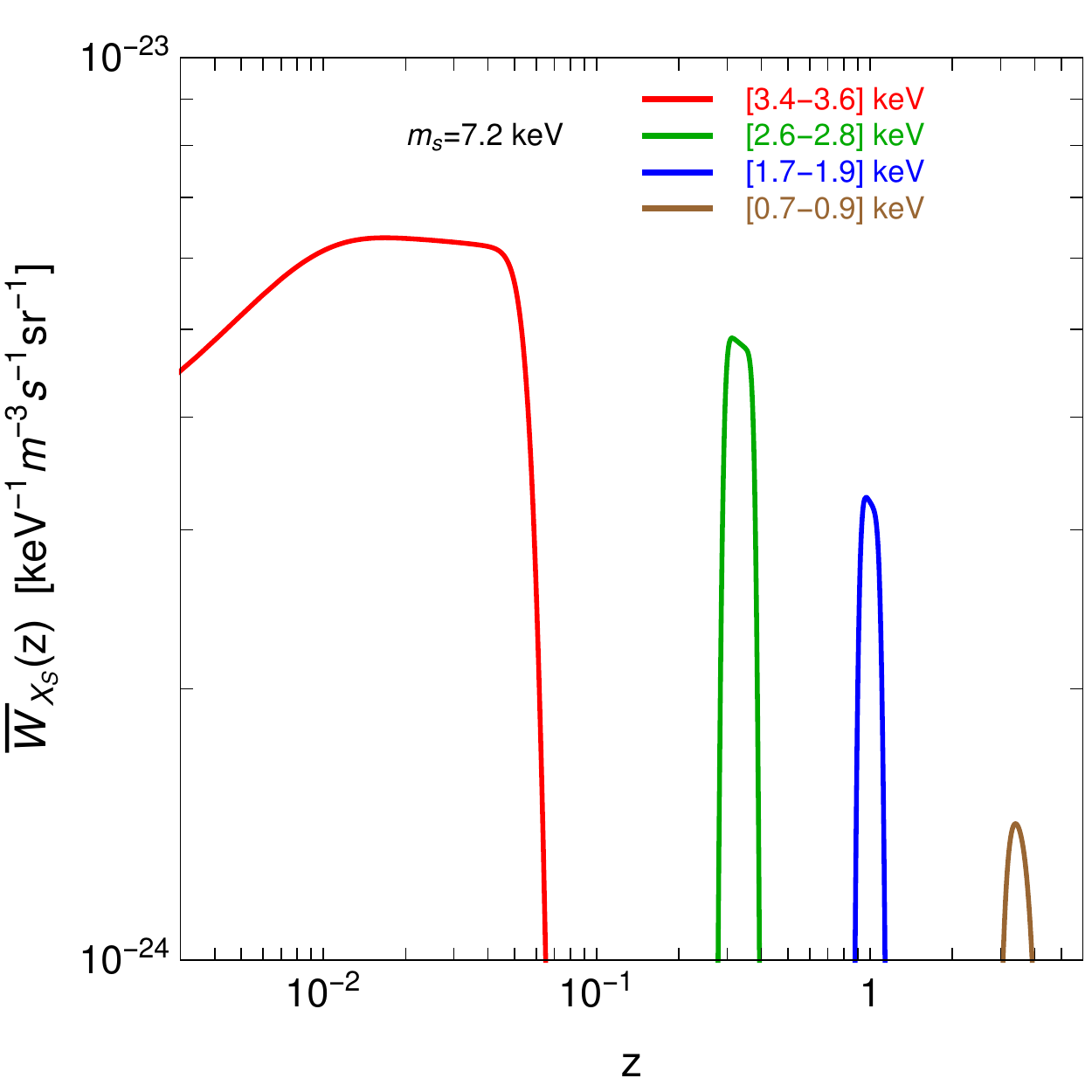}
\caption{Left: window functions of the catalogs of galaxies considered in the analysis. All the window functions are normalized to unit, $\int dz\,dN_g/dz=1.$ Right: average window function of sterile neutrino in different energy bins, i.e.,  $\bar{W}_{X_S}(z)\equiv1/\Delta E\,\int_{\Delta E} dE \, W_{X_S}(E,z).$ In this example, the DM mass is $7.2$ keV and $\sin^2(2\theta)=3.4\times10^{-11}.$
}
\label{fig:Wfunctions}
\end{center}
\end{figure}

In our modeling of the background we do not include emission lines. They can be given by several atomic processes, but the exact strength, size, dependence on the type of astrophysical object, and redshift scaling are typically quite unknown (for an example, see the debate on the interpretation of the 3.5 keV line \cite{2014arXiv1409.4143B}). For sufficiently large energy bins, these lines are not expected to significantly alter our estimate of the background, and therefore of the covariance entering the computation of the projected bounds (but see comments below on the Athena X-IFU case). On the other hand, in the case of detection of a line feature, clearly, there will be some degeneracy between the interpretation in terms of sterile neutrino decay and an atomic line. However, since the latter is associated to astrophysical processes, it typically has quite different redshift behaviour and width with respect to the DM-induced line. These two handles can help in disentangling between the two cases.

\begin{figure}[t]
\begin{center}
\includegraphics[width= 0.48 \textwidth]{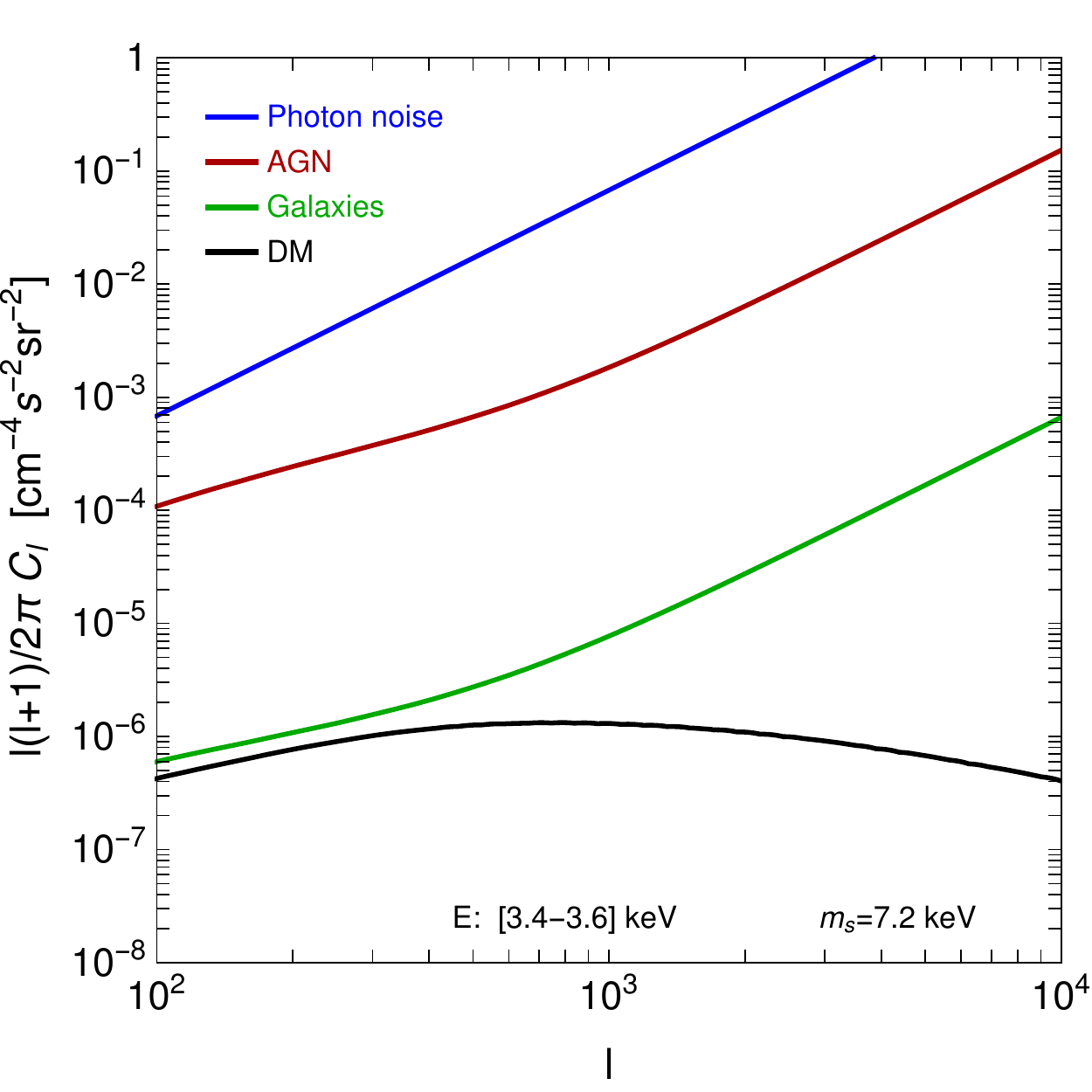}
\hspace{0.27cm}
\includegraphics[width= 0.48 \textwidth]{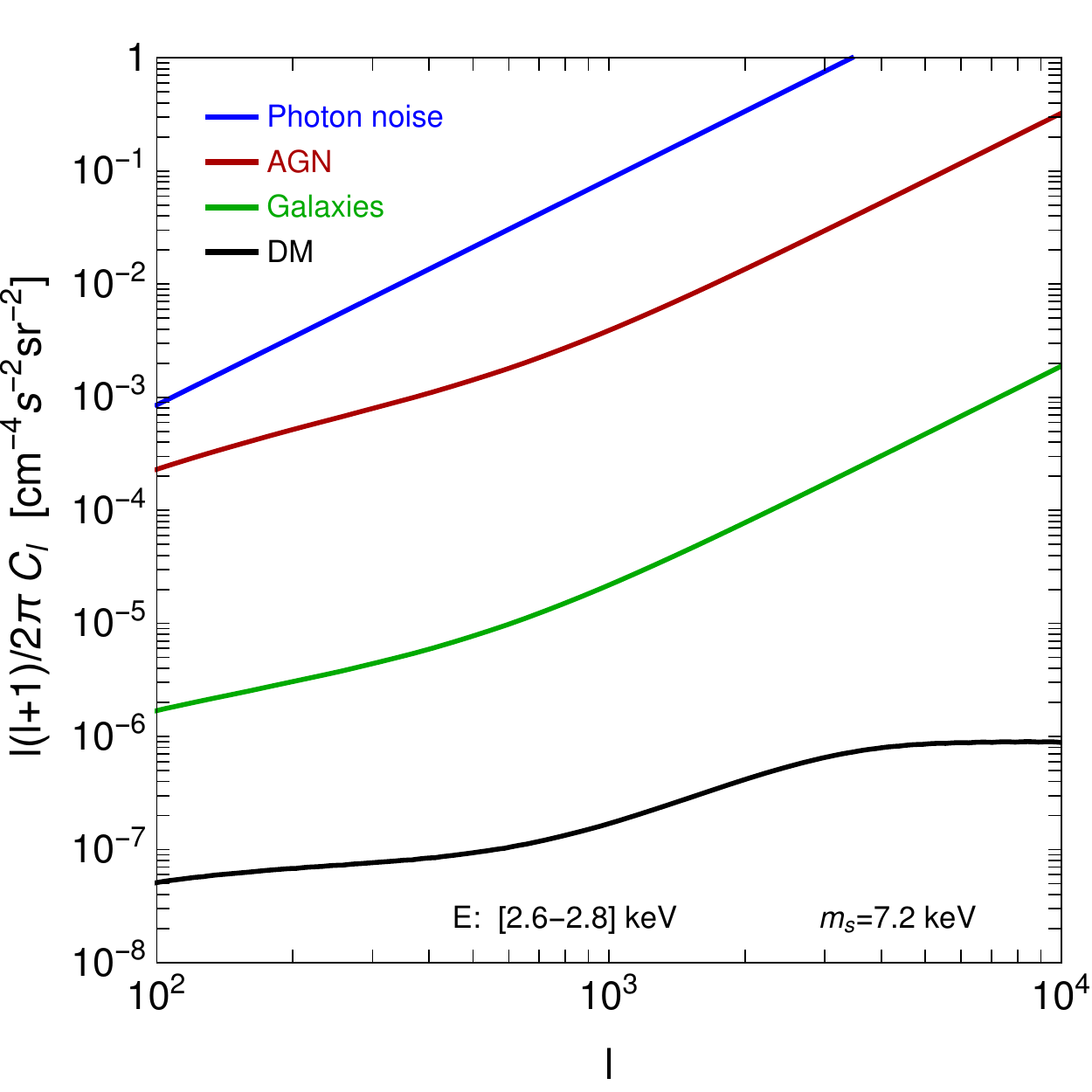}
\caption{Auto-correlation APS of the different X-ray extragalactic emitters as predicted for eROSITA. The DM mass and $\sin^2(2\theta)$ are as in~\Fig{fig:Wfunctions}. Left and right plots show different energy bins, $[3.4,3.6]$ and $[2.6,2.8]$ keV, respectively.
The blue line shows the photon noise, described in \Sec{sec:res}.}
\label{fig:Auto}
\end{center}
\end{figure}

\begin{figure}[t]
\begin{center}
\includegraphics[width= 0.48 \textwidth]{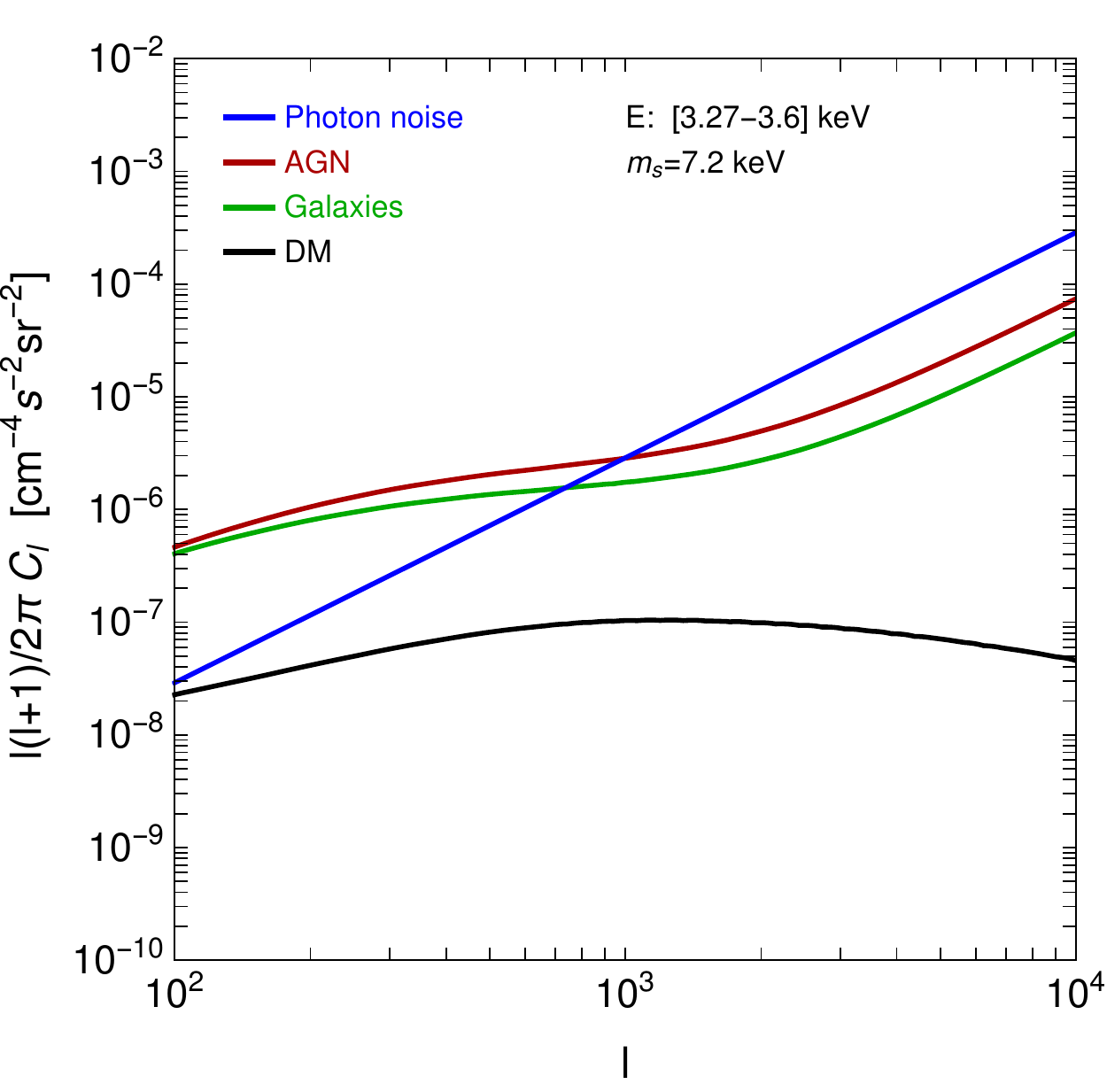}
\hspace{0.27cm}
\includegraphics[width= 0.48 \textwidth]{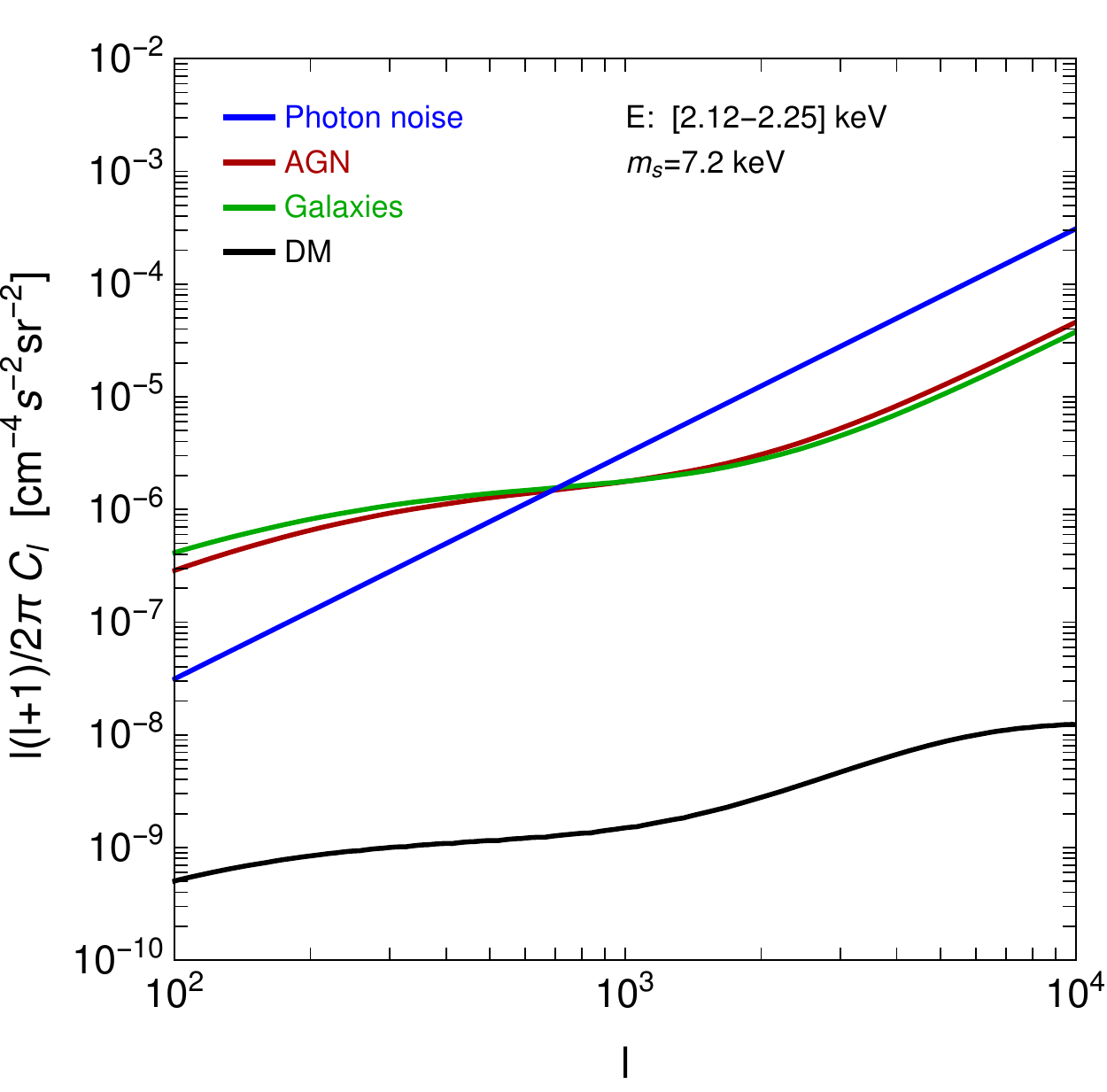}
\caption{Auto-correlation APS of the different X-ray extragalactic emitters as predicted for Athena WFI. 
The DM mass is $7.2$ keV and $\sin^2(2\theta)=1.5\times10^{-12}$.
Left and right plots show different energy bins, $[3.27,3.6]$ and $[2.12,2.25]$ keV, respectively.
The blue line shows the photon noise, described in \Sec{sec:res}.}
\label{fig:AutoAthena}
\end{center}
\end{figure}

For what concerns the description of the catalogs of galaxies, the window function is simply provided by the redshift distribution of the objects, $dN_g/dz$.
 More precisely, $W_{g}(z)= H(z)/c\,dN_g/dz$ such that $\int \de\chi W_{g}(\chi)=1$ for a redshift distribution $dN_g/dz$ normalized to unity. 
 The form of $dN_g/dz$ for the different catalogs adopted in our work is shown in the left panel of \Fig{fig:Wfunctions}.
 
 On the right panel of \Fig{fig:Wfunctions} we show the window function of DM, for an example mass of 7.2 keV and in a few different energy bins. Note that they correspond to a given redshift bins. Therefore the cross correlation of a given energy bin selects a specific redshift slice in the galaxy distribution, something which is not true for the astrophysical sources (having a continuum energy spectrum) and is a key point of the line intensity mapping strategy. 

\begin{figure}[t]
\begin{center}
\includegraphics[width= 0.48 \textwidth]{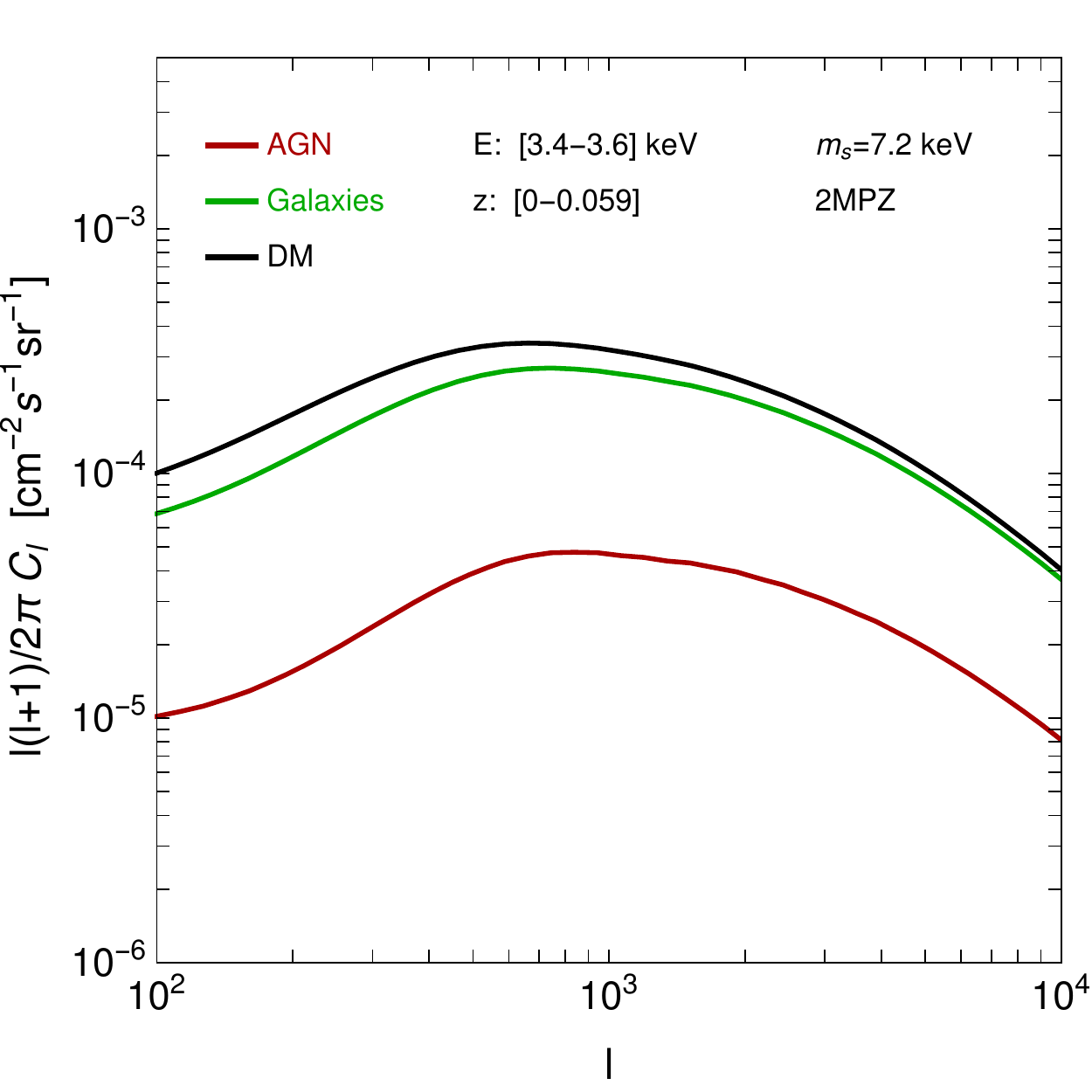}
\hspace{0.27cm}
\includegraphics[width= 0.48 \textwidth]{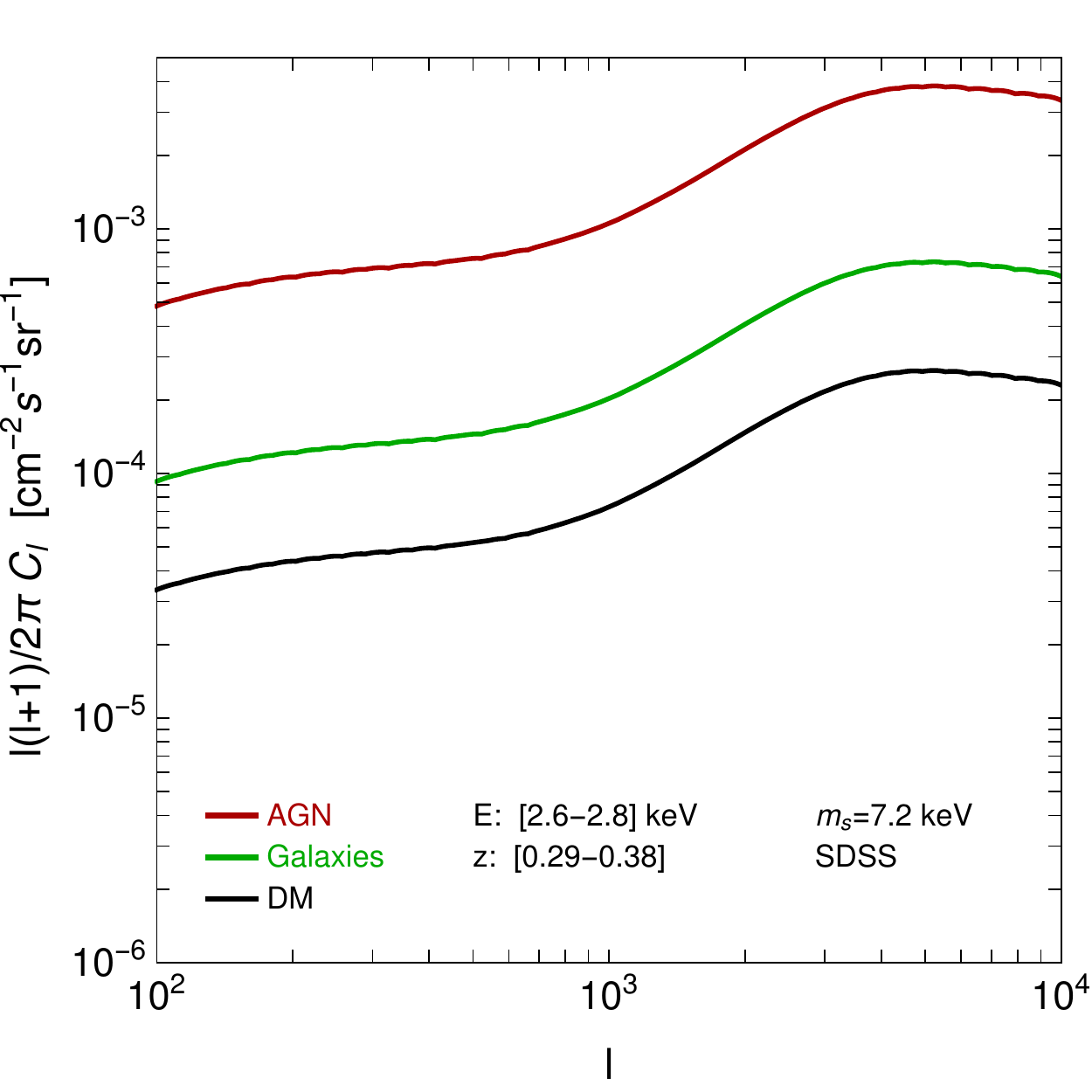}
\caption{Same as in~\Fig{fig:Auto} but for the cross-correlation APS. The corresponding redshift bins are selected from the relation $z = m_S/(2\,E)-1.$
}
\label{fig:Cross}
\end{center}
\end{figure}

\begin{figure}[t]
\begin{center}
\includegraphics[width= 0.48 \textwidth]{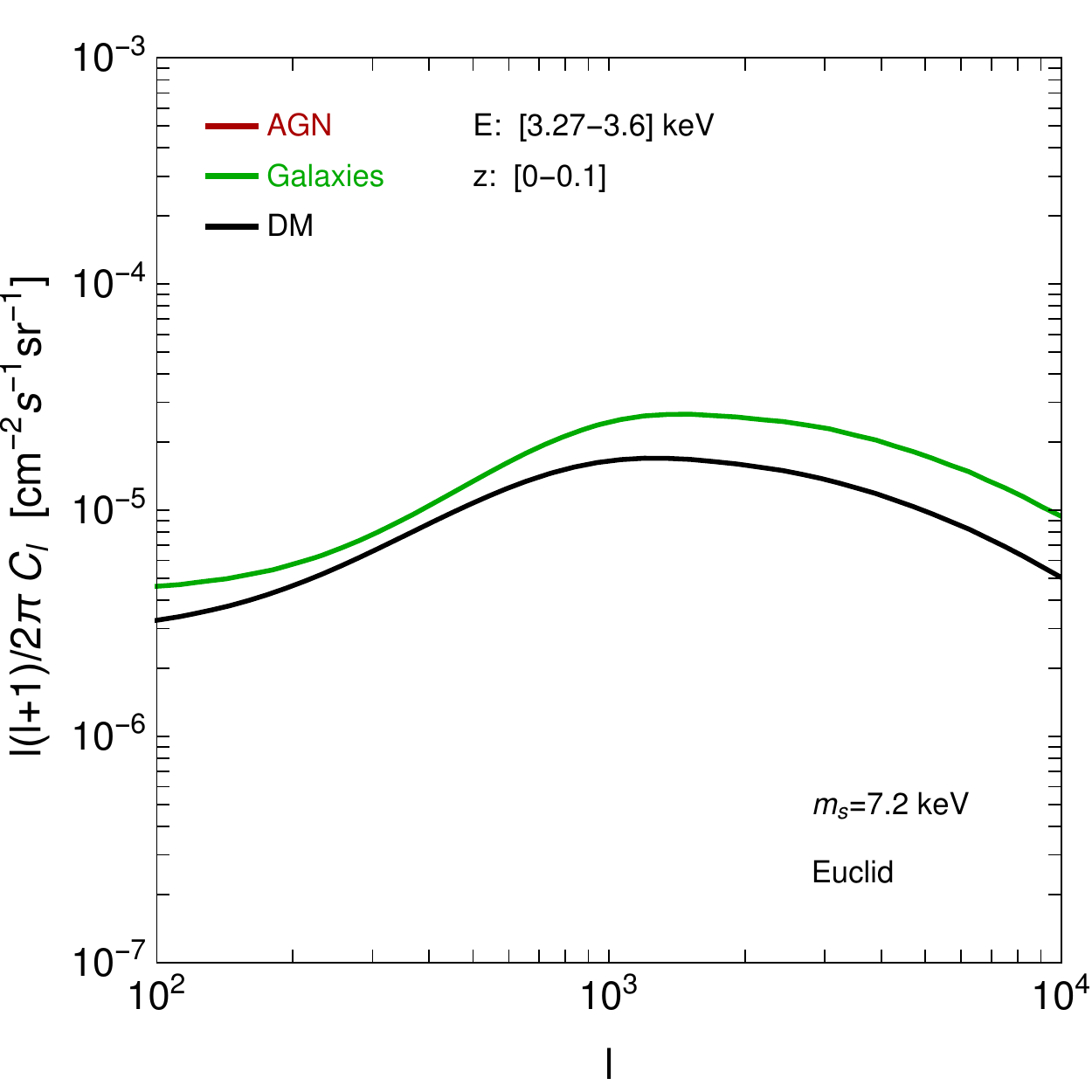}
\hspace{0.27cm}
\includegraphics[width= 0.48 \textwidth]{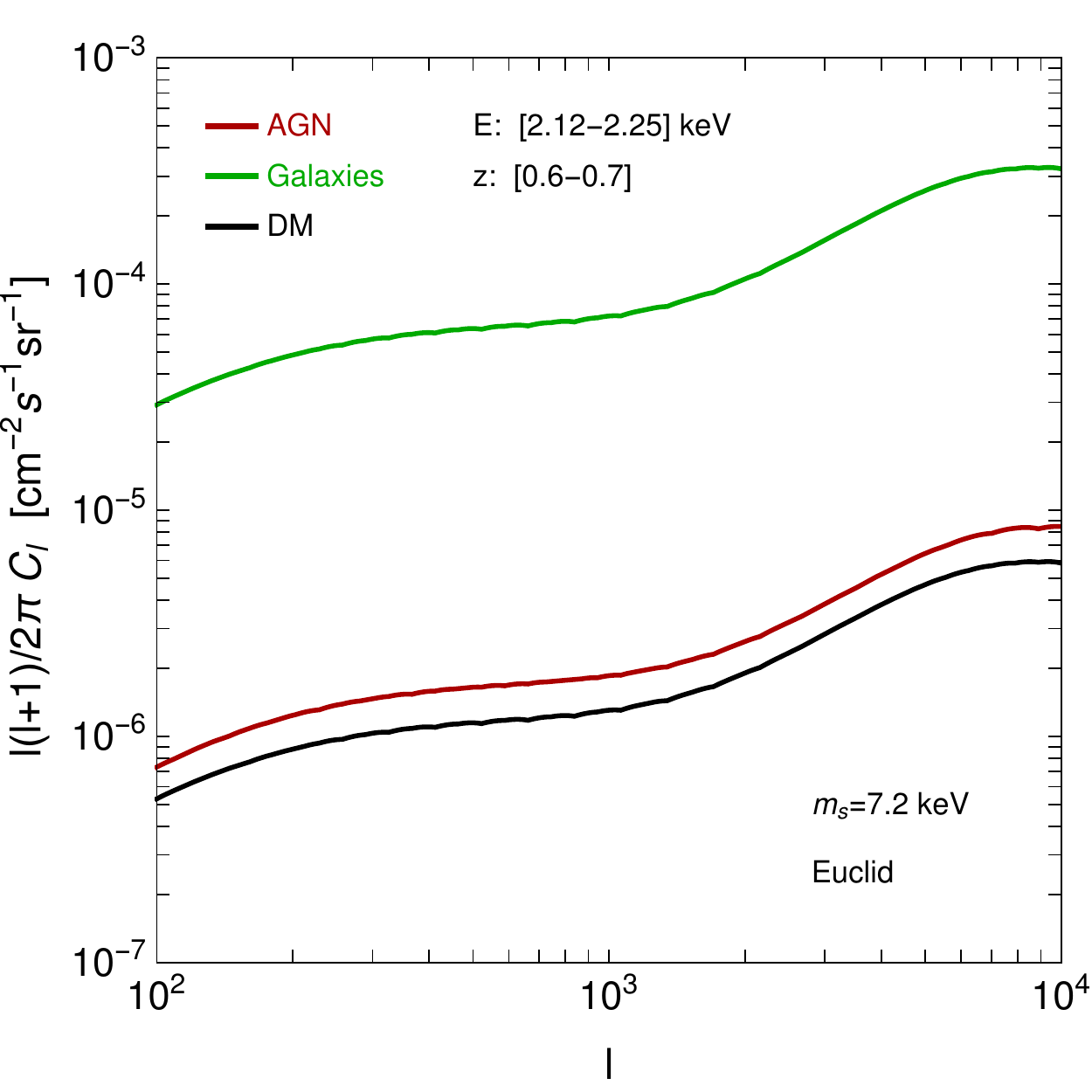}
\caption{Same as in~\Fig{fig:AutoAthena} but for the cross-correlation APS. The corresponding redshift bins are selected from the relation $z = m_S/(2\,E)-1.$
}
\label{fig:CrossAthena}
\end{center}
\end{figure}

The angular power spectrum (APS) of the cross correlation between a map of X-rays in the $a$th energy bin and a catalog of galaxies in the $r$th redshift bin can be computed as:
\be
C_\ell^{ar} = \int_{\Delta z_r} \de z\,\int_{\Delta E_a}\de E\,\frac{c}{H(z)}\frac{W_\textrm{\rm X}(E,z)W_{\rm g}(z)}{\chi(z)^2}
P_{\rm X,g}\!\!\left[k=\frac{\ell}{\chi(z)},z\right],
\label{eq:clcross}
\ee
where $P_{\rm X,g}$ is the three-dimensional cross power spectrum between a given X-ray population and a given galaxy catalog. It is a function of both redshift and modulus of the physical wavenumber $k$. In the Limber approximation~\cite{1953ApJ...117..134L}, $k$ and the angular multipole $\ell$ are linked by $k=\ell/\chi(z)$. This approximation is valid for $\ell\gg1$, i.e., in the range considered in the present work.

In the case of X-ray auto correlation (relevant to compute the covariance of the cross correlation), the expression of the APS reduces to:
\be
C_\ell^{ab} = \int \de z\,\int_{\Delta E_a}\de E\, \int_{\Delta E_b}\de E'\,\frac{c}{H(z)}\frac{W_\textrm{\rm X}(E,z)W_\textrm{\rm X}(E',z)}{\chi(z)^2}
P_{\rm X}\!\!\left[k=\frac{\ell}{\chi(z)},z\right],
\label{eq:clauto}
\ee

We will compute the 3D power spectrum using the halo model formalism (for a review, see, e.g., Ref.~\cite{Cooray:2002dia}), where the power spectrum is described in terms of the sum of the one-halo ($P^{1h}$) and two-halo ($P^{2h}$) components.
Their generic expression is given by (see, e.g., Ref.~\cite{Fornengo:2013rga}):
\bea
 P_{ij}^{1h}(k) &=& \int dM\ \frac{dn}{dM} \hat f_i^\ast(k|M)\,\hat f_j(k|M) \label{eq:PShalo}
 \\
 P_{ij}^{2h}(k) &=& \left[\int dM\,\frac{dn}{dM} b_i(M) \hat f_i^\ast(k|M) \right] \
                \left[\int dM\,\frac{dn}{dM} b_j(M) \hat f_j(k|M) \right]\,P^{\rm lin}(k)\nonumber
\eea
where $M$ is the halo mass, $dn/dM$ is the halo mass function, $\hat f$ is the Fourier transforms of the field under consideration, $b$ is the bias of the source with respect to matter and $P^{\rm lin}$ is the linear matter power spectrum.
The expressions for the various cases considered in our work are detailed in the Appendix.

Let us note here that we will be considering the power spectrum of cold DM, even though the sterile neutrino can be a warm DM candidate. The deviation of its power spectrum from a pure cold DM power spectrum strongly depends on its production mechanism. On the other hand, luckily, this does not impact our results, since the signal we are considering is mostly produced in massive halos, where differences between cold and warm scenarios are negligible (see also Ref.~\cite{Zandanel:2015xca}).

    Examples of the auto- and cross-APS are shown in Figs.~\ref{fig:Auto}-\ref{fig:CrossAthena}. In \Fig{fig:Auto} and \Fig{fig:Cross}, we consider the eROSITA experimental setup (described in the following), a sterile neutrino with mass $m_S=7.2$ keV, and two energy bins $[3.4,3.6]$ and $[2.6,2.8]$ keV. For the cross-correlation plots we select the redshift bins providing the DM signal ($[0,0.059]$ for cross correlation with the $[3.4,3.6]$ keV energy bin, and $[0.29,0.38]$ for the $[2.6,2.8]$ keV energy bin) and consequently the two catalogs whose galaxy distribution is peaked in such bins, that are, respectively, 2MPZ and SDSS.
For the same DM mass, in \Fig{fig:AutoAthena} and \Fig{fig:CrossAthena} we show the auto- and cross-APS as predicted for Athena. We cross correlate the X-ray signals with the Euclid survey of galaxies. We select a low redshifit bin $[0,0.1]$ and  one closer to the peak of the galaxy distribution, $[0.6,0.7]$. The energy bins are those providing the DM signal, i.e. $[3.27-3.6]$ keV and $[2.12-2.25]$ keV.

Note that, while the auto correlation is largely dominated by AGN (and photon noise), the DM signal becomes important in the cross correlation, especially at low redshift. This highlights the importance of a tomographic approach involving DM tracers at different redshift in order to distentangle the DM cosmological X-ray emission from astrophysical contributions.

\begin{table}[t]
 \small
  \centering
 \renewcommand\arraystretch{2.2}
\begin{tabular}{|c||c|c|c|}
\hline
 & eROSITA & Athena WFI & Athena X-IFU  \\ \hline \hline
Energy range [keV] & 0.3-10 & 0.1-12 & 0.3-12 \\ \hline 
$A_{\rm eff}$ at 3 keV [m$^2$]& 0.03 & 0.79 & 0.68   \\ \hline 
$\Omega_{FoV}$ [deg$^2$] & 0.66 & 0.69 & 0.014  \\ \hline 
HEW [arcsec] & 28 & 5 & 5  \\ \hline 
\makecell{Spectral resolution (FWHM) \\ at 7 keV [eV] }& 138 & 138 & 2.5  \\ \hline 
\makecell{$F_{\rm sens}$ \\ $$[erg cm$^{-2}$ s$^{-1}$]} & $1.1\times10^{-14}$ & $2.4\times10^{-17}$ & $2.4\times10^{-17}$  \\
\hline
\makecell{Particle bkg \\$$[counts keV$^{-1}$ s$^{-1}$ sr$^{-1}$]}& $1.2 \times 10^3$ & $1.2 \times 10^3$ & $5.8 \times 10^3$   
\\ \hline 
\end{tabular} 
\caption{Key performances of the three X-rays instruments considered in the analysis.
The rows report the energy range, the effective area at $3$ keV, the field of view, the angular resolution (Half Energy Width), the energy resolution (at $7$ keV), the source sensitivity in the $0.5-2$ keV band (considering observations in survey mode for eROSITA, while 450 ks exposure in the Athena cases) and the rate of particle background.}
\label{tab:experiments}
\end{table}

\section{Experiments}
\label{sec:exp}

The eROSITA instrument, in orbit since July 2019, is going to perform a deep survey of the entire sky in the $0.3-10$ keV energy range, with unprecedented angular and energy resolution.
It is therefore ideally placed to search for the cosmological DM signal described in the previous Section.
In \Tab{tab:experiments} we summarize its performances for those parameters more relevant to our analysis.
The energy-dependent effective area and spectral resolution are taken from Refs.~\cite{Merloni:2012uf,eRosita,Merloni}. We consider an observational time of four years, corresponding to the duration of the all-sky survey program. For our sensitivity forecasts we focus on a fraction of sky $f_{\rm sky}^X=0.8,$ after masking the Galactic plane and resolved sources. 
As explained in~\Sec{sec:res}, we need to model the total intensity received by instrument, which is the sum of the extragalactic emission in Eq.~\ref{eq:intensity}, the Galactic contribution and the instrumental background.
The Galactic foreground, modeled as in Section 4.2 of Ref.~\cite{Merloni:2012uf}, is relevant only at energies $\lesssim$ $1$ keV. The particle background is generated by the interactions of particles, mostly protons and secondary electrons, with the instrument. The expected rate is 1151 counts keV$^{-1}$ s$^{-1}$ sr$^{-1}$~\cite{Merloni:2012uf}.

The Athena X-ray observatory is planned to be launched in early 2030s. The experiment will host two instruments, the Wide Field Imager (WFI) and the X-ray Integral Field Unit (X-IFU). 
The performances of the instruments, taken from Ref.~\cite{Nandra:2013shg}, are summarized in \Tab{tab:experiments}.

Athena WFI will have a field of view similar to eROSITA, with a dramatically improved effective area (taken from Ref.~\cite{Barret:2013bna}).
Athena X-IFU will deliver in-depth spectroscopic observations, thanks to a superior spectral resolution ($2.5$ eV for $E<7$ keV and $E/\Delta E=2800$ for $E>7$ keV, see Ref.~\cite{xIFU}), but over a more limited field of view.

For the sake of definiteness, we consider the average exposure for each field observed by Athena to be 450 ks.
Taking a total observing time of 5 years, corresponding to the integrated nominal mission lifetime, this leads to a number of observed fields of $5 {\rm yr} / 450 {\rm ks} \simeq 350$.
The sky coverage is thus computed as $f_{\rm sky}^X=350\, \Omega_{FOV}$, resulting to $f_{\rm sky}^{WFI}=5.9\times 10^{-3}$ and 
$f_{\rm sky}^{X-IFU}=1.2\times 10^{-4}$.

The rate of particle background is obtained following Ref.~\cite{Lotti:2014kda}.
\medskip

\begin{table}[t]
 \small
  \centering
 \renewcommand\arraystretch{2.2}
\begin{tabular}{|c||c|c|c|c|c|c|}
\hline
 & 2MPZ & SDSS & DES & DESI & Euclid &LSST\\ \hline \hline
\# of galaxies & $6.7\times 10^5$ & $1.5\times 10^7$ & $2\times 10^8$ & $3.5\times 10^7$ & $2\times 10^9$& $3.6\times 10^9$\\ \hline 
Sky coverage & 0.66 & 0.26 & 0.12 & 0.34 & 0.36& 0.49\\ \hline 
$z$-range & 0-0.08 & 0.08-0.8 & 0.08-2 & 0-2 & 0-2& 0-3\\ \hline 
\# of $z$-bins & 1 & 4 & 6 & 132$^*$ & 7& 8\\ \hline 
Reference & \cite{Bilicki14} & \cite{SDSS2016} & \cite{Abbott:2005bi} & \cite{Aghamousa:2016zmz} & \cite{Euclid2011}&\cite{Abell:2009aa}\\ \hline 
\end{tabular} 
\caption{Properties of the galaxy catalogs considered in this work. First line reports the total number of objects in the catalog, second line is the fraction of sky covered by the survey (after masking), third and fourth lines show the redshift range and number of bins we considered in our analysis for each catalog.\newline
$^*$The number of $z$-bins quoted for DESI is in the case of cross correlation with Athena data, see text for the case of eROSITA.}
\label{tab:catalogs}
\end{table}

In order to maximize the cross-correlation signal, we selected the deepest and widest, in terms of number of objects and sky coverage, surveys of galaxies at mid-low redshift of the near past/future.
Photometric surveys have the advantage of detecting a larger number of galaxies, while spectroscopic surveys are superior in the search for a spectral line. We consider both cases.

Concerning available data, we selected 2MPZ~\cite{Bilicki14} at low-$z$ and SDSS~\cite{SDSS2016} at mid-$z$. At the time of writing, the largest photometric survey is DES~\cite{Abbott:2005bi}, which has completed data-taking and is finalizing the data-reduction.
The spectroscopic survey DESI~\cite{Aghamousa:2016zmz} just started observations and will have a timeline similar to eROSITA.
In a few years, two wide-field surveys, Euclid~\cite{Euclid2011} and LSST~\cite{Abell:2009aa}, will reach the milestone of observing billions of galaxies, thanks to their deep sensitivity (35 gal/arcmin$^2$ and 50 gal/arcmin$^2$, respectively).
They are expected to complete observations before/around the launch of Athena satellite.

In \Tab{tab:catalogs}, we list the catalogs considered in our analysis and report the relevant properties entering the computation of the cross-correlation signal and covariance matrix.
The redshift distribution of the galaxies in each catalog is shown in \Fig{fig:Wfunctions}.

\begin{figure}[t]
\begin{center}
\includegraphics[width= 0.48 \textwidth]{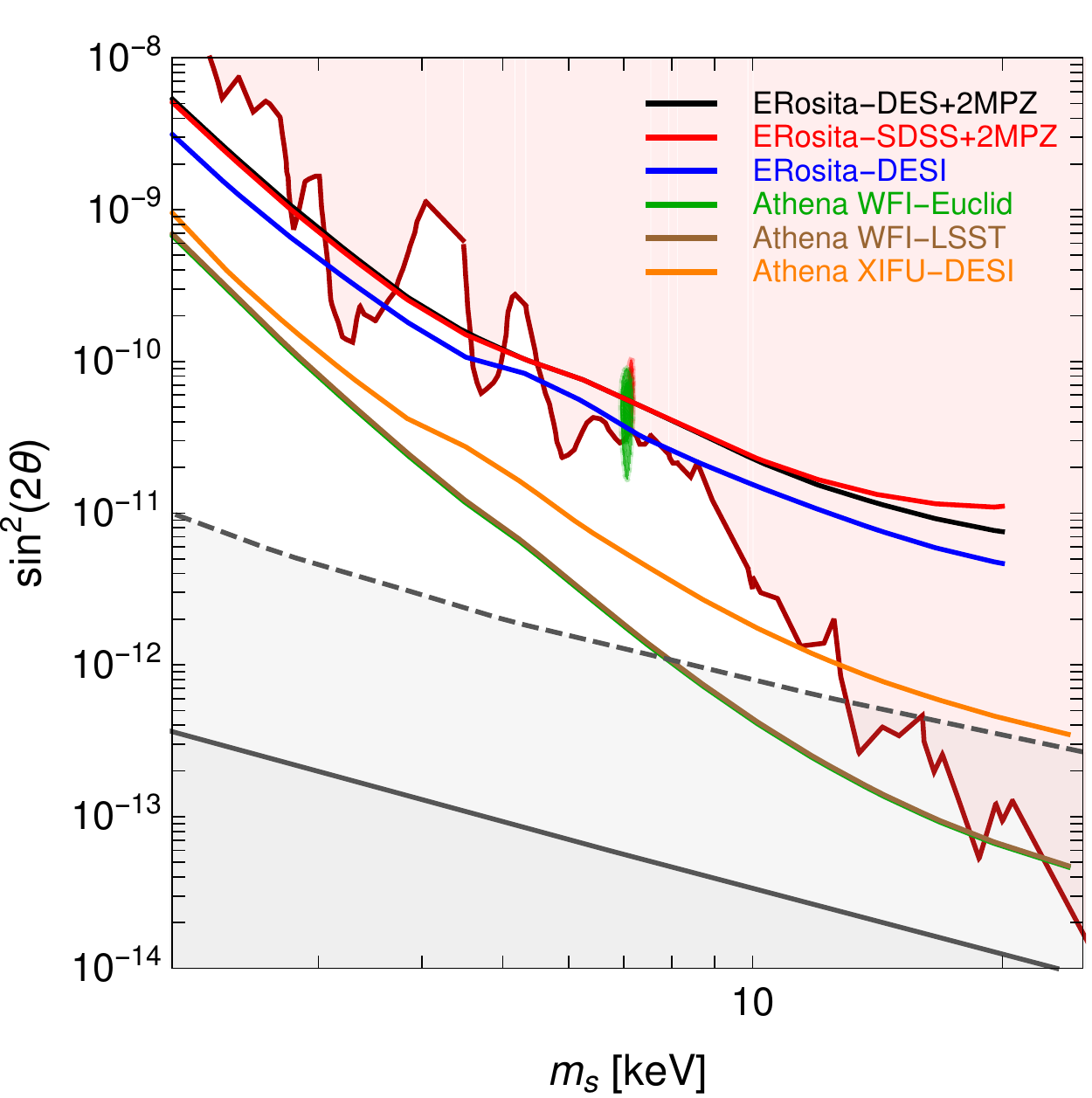}
\hspace{0.27cm}
\includegraphics[width= 0.48 \textwidth]{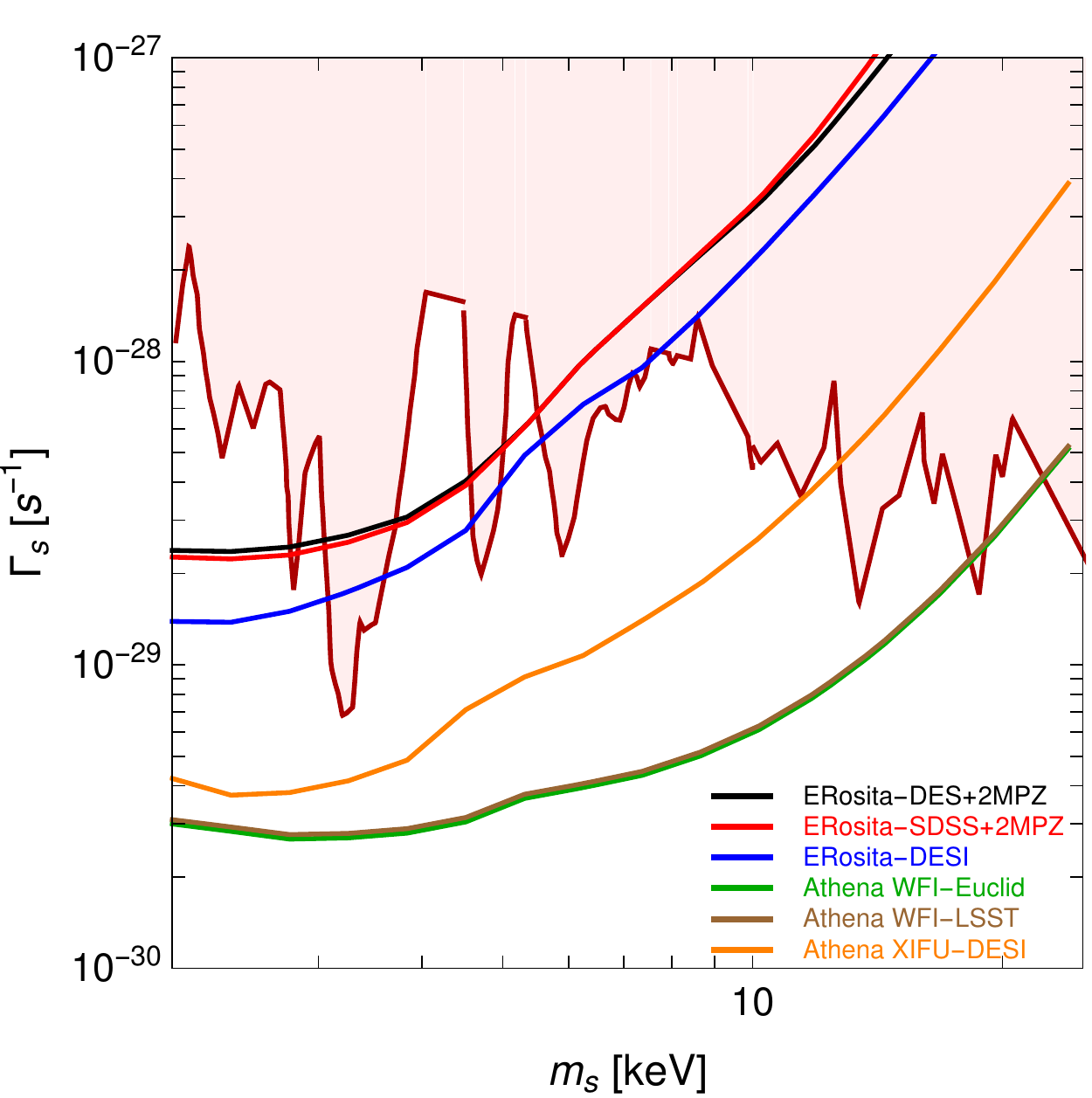}
\caption{Projected 95\% C.L. bounds for the different cross-correlation analyses. The left plot presents the results in the sterile neutrino $m_S$ vs $\sin^2(2\theta)$ plane. 
The light red region is excluded by current X-ray observations~\cite{Abazajian:2017tcc,Roach:2019ctw}.
The red and green contours are the 2-$\sigma$ regions for the $3.55$ keV line excess respectively from the MOS stacked clusters~\cite{Bulbul:2014sua} and M31~\cite{Boyarsky:2014jta}.
For the case of resonant production, the solid (dashed) gray line shows where sterile neutrino accounts for all the DM for a lepton asymmetry $L=7\cdot10^{-5}$  ($L=2.5\cdot 10^{-3}$, the maximum value allowed by BBN~\cite{Boyarsky:2009ix}). Below these lines sterile neutrino is a subdominant DM component (for these choices of $L$ and considering resonant production only).
The right plot presents the projected bounds and current X-ray constraints in the plane DM decay rate versus DM mass. 
}
\label{fig:Sensitivities}
\end{center}
\end{figure}

\section{Results}
\label{sec:res}
In \Sec{sec:models}, we depicted how to compute the auto- and cross-correlation APS, and in \Sec{sec:exp} we summarized the experimental quantities needed for their estimates.
Now we describe how bounds on the sterile neutrino properties are derived from such forecasts and we discuss our findings.

In order to define a statistical significance, we need first to introduce the covariance matrix associated to the cross-correlation APS. In the Gaussian approximation, it can be written as:

\be
\Gamma_{ar\ell,br^\prime\ell^\prime} = \frac{ \delta^{\rm K}_{\ell\ell^\prime}}{(2\ell+1)\Delta \ell f_{\rm sky} }\left[C_\ell^{ar}C_{\ell^\prime}^{br^\prime}+ \big(C_\ell^{ab} + C_\nn^{ab}\delta^{\rm K}_{ab}\big) \big(C_{\ell^\prime}^{rr^\prime}+ C_\nn^{rr^\prime}\big) \delta^{\rm K}_{r r^\prime} \right]\;,\label{eq:covCl}
\ee
where again $a$ and $b$ label energy bins of X-ray measurements, and $r$ and $r^\prime$ label redshift bins in the catalogs.
The analysis is performed over 40 multipole bins in the $\ell$-range $[10^2,10^4]$, with even logarithmic spacing. The size of the bins is pretty large, and this further justifies the Gaussian approximation of the covariance.

The auto- and cross-correlation APS entering in Eq.~\ref{eq:covCl} are described in \Sec{sec:models}. The photon noise term is given by $C_\nn^a=4\pi f_{\rm sky}^X\,W_\ell^{-2}\langle I_X^a \rangle^2/N_X^a$, where $\langle I_X^a \rangle$ is the sky-averaged X-ray intensity observed by the telescope in the $a$th energy bin and given by the sum of the extragalactic contributions computed from Eq.~\ref{eq:intensity} plus the Galactic emission plus the particle background, with the latter two described in \Sec{sec:exp}; the number of observed photons is $N_X^a=\langle I_X^a \,A_{\rm eff}^a \rangle\,t_{\rm obs}\,\Omega_{FoV}$; $W_\ell$ is the beam window function whose effect is however negligible at $\ell<10^4$ given the angular resolution (HEW) reported in \Tab{tab:experiments}.
%is approximated with a Gaussian $W_\ell=\exp{(-\sigma_{\rm PSF}\ell^2/2)}$ (its effect is in any case marginal at $\ell<10^4$).
The shot noise term for galaxies is given by $C_\nn^r=4\pi f_{\rm sky}^g/N_g^r$, where the number of galaxies in the $r$th redshift bin is computed from the distribution shown in \Fig{fig:Wfunctions}. Here we neglect the beam window function since again the typical angular resolution is significantly smaller the smallest scale considered in our analysis.

The fraction of sky in Eq.~\ref{eq:covCl} is taken to be the smallest between X-ray and galaxy surveys, i.e., $f_{\rm sky}={\rm min}[f_{\rm sky}^X,f_{\rm sky}^g]$. In the case of eROSITA, the galaxy surveys have a smaller sky coverage, while we have the opposite picture in the case of Athena.

The upper limits on the sterile neutrino parameter space are derived at 95\% C.L. by requiring $\chi^2=2.71$ with the estimator assumed to follow a $\chi^2$ distribution with one dof (i.e., $\sin^2(2\theta)$ for a given $m_S$)  defined as:
\be
\chi^2=\sum_{a,b,\ell} C_{\ell,S}^{ar}\Gamma_{ar\ell,br^\prime\ell}^{-1}C_{\ell,S}^{br^\prime}\;.
\label{eq:chi2}
\ee
There are two things to note. First, only the cross-correlation term involving sterile neutrinos is present in the signal part of Eq.~\ref{eq:chi2}. This is because we assume to be able to extract the background associated to X-ray emitting AGN and galaxies to a good precision, and so we neglect model uncertainties in the astrophysical components. This assumption is well justified in the case of spectroscopic surveys, where there is a huge number of cross energy-redshift terms not entering in the signal of Eq.~\ref{eq:chi2}, and that can be used to fit a continuum term. In the case of photometric surveys, it has instead to be checked with data at hands whether this assumption is fully valid.
As stated in the Introduction, this is the reason why a spectroscopic approach that fully exploits the line intensity mapping can be considered more robust.
The second crucial point to note in Eq.~\ref{eq:chi2} is that the sum does not run over the redshift bins. This is because for a given energy bin we select the corresponding redshift bin from the relation $z = m_S/(2\,E)-1$. As already mentioned above, the DM signal in the other redshift bins is predicted to be null.

The contamination from emission lines deserves a separate discussion. The emission from the Galaxy or from the instrument only affects the covariance estimate (i.e., it is not correlated with extragalactic surveys). In the case of eROSITA and Athena WFI this occurs at a negligible level (for the energy range of interest), due to the smearing associated to the energy resolution, included in the model we are adopting \cite{Merloni:2012uf}. For Athena X-IFU, these lines can be sizable in a handful of energy bins. However, they can be just removed from the analysis without compromising the sensitivity.

On the other hand, emission lines from extragalactic sources can be more degenerate with the sterile neutrino signal, if occurring at the same energy, and can have a larger impact on the error estimate (in particular in the case of Athena X-IFU).
The degeneracy might be broken looking at the different properties of two lines, as mentioned in \Sec{sec:models}, but still the sensitivity on a few DM masses can be affected and the final exclusion bound would result to be more jagged than our ``smooth'' projected curve, which is intended to show the attainable level with a good cleaning of line contamination.

Results obtained with the assumptions and procedure outlined above are shown in \Fig{fig:Sensitivities}, which reports the 95\% C.L. bounds in the plane $\sin^2(2\theta)$ versus $m_S$.

\subsection{eROSITA}
\label{sec:rosita}
We analyze the cross-correlation of the X-ray sky from eROSITA (which recently started its operation, as mentioned in \Sec{sec:exp}) with past/ongoing experiments.
In particular, the red, black and blue lines in \Fig{fig:Sensitivities} show the bounds from the cross correlation with SDSS+2MPZ, DES+2MPZ, and DESI, respectively.
In the case of photometric surveys, we take 2MPZ as the catalog of reference at low-$z$ considering a bin $z=[0,0.08]$, and four (six) additional bins at $z>0.08$ referring to SDSS (DES) data.
Despite DES observes a significantly larger number of galaxies than SDSS, it covers a smaller portion of the sky (see \Tab{tab:catalogs}), and the associated bounds come out to be comparable. In other words, DES has a lower noise term but higher cosmic variance contribution with respect to SDSS. This means to have a lower $C_\nn$ but also lower $f_{\rm sky}$ in Eq.~\ref{eq:covCl} with the two effects (accidentally) compensating each other in the computation of the bound.

It is interesting to note that the case of cross correlation with the DESI catalog is more constraining than SDSS despite having similar number of galaxies and sky coverage. This is because in the search for a line, clearly, the closer the size of the energy/redshift bin is with respect to the line width the higher is the signal to noise ratio.
On the other hand, the increase in sensitivity is not dramatic because the spectral resolution of eROSITA prevents to have very narrow bins. Namely their size ranges from $E/\sigma_{FWHM}=59$ to $E/\sigma_{FWHM}=6$ for $E=10$ keV and $E=0.3$ keV respectively. 

In \Fig{fig:Sensitivities}, we show also the preferred regions for the possible excess found in a few X-ray observations and interpreted as a potential signature of sterile neutrino. The regions are taken from Refs.~\cite{Bulbul:2014sua,Boyarsky:2014jta}.
Note that with the technique proposed in this paper, it will be possible to test this interpretation with the data acquired by eROSITA in the forthcoming years.

\subsection{Athena WFI and X-IFU}
\label{sec:athena}
In the epoch of Athena observations, the data from the surveys of DESI, Euclid, and LSST will be available.
We show the corresponding bounds on the sterile neutrino properties from the cross-correlation analysis with orange, green, and brown lines, respectively. 

The cases of photometric surveys (Euclid and LSST) cross correlated with Athena WFI provide an improvement of more than one order of magnitude with respect to the case of eROSITA cross correlated with DES. This is due to improvements on both sides: larger $f_{\rm sky}$ and number of galaxies on the catalog side, and improved effective area of Athena versus eROSITA.
In particular the latter allows to significantly reduce the photon noise term $C_\nn^a$ in Eq.~\ref{eq:covCl}, overcoming the reduction in the sky coverage of ATHENA with respect to eROSITA.
The derived bound has the capability to close the allowed window for a resonantly produced sterile neutrino parameter for $m_S>7$ keV (and lepton asymmetry of $L\lesssim7\cdot10^{-5}$).
Despite the number of galaxies in LSST is larger with respect to Euclid, the bounds are similar. This is because those galaxies are added at high redshift, where the contribution of the signal to the $\chi^2$ is very small.

Athena X-IFU is potentially a game-changer, in the sense that its superior spectral resolution makes the line intensity mapping fully attainable. On the other hand, the reduced $\Omega_{FoV}$ with respect to the WFI detector (fifty times smaller, see \Tab{tab:experiments}), leads to a reduced sky coverage (for the same flux sensitivity). In order to compensate for this, we should consider (around) fifty times more redshift bins than in the WFI case.
This is in principle possible at high X-ray energy. On the other hand, such a thinner binning would bring us to consider a full 3D correlation instead of the 2D case analysed throughout the paper, in order to include effects like redshift space distorsion. We leave this improvement for future developments, while here we set the spectral resolution of the correlation DESI-Athena X-IFU to $R=120$ (corresponding to $\Delta z\sim 40$ Mpc), which also corresponds to the X-IFU resolution at low energy. In this way, the DM velocity dispersion in halos can be safely neglected.
Note that this bound, contrary to the photometric case, does not benefit from an improvement on the galaxy survey side since we consider DESI as for the eROSITA bound (whilst we move from DES to Euclid/LSST in the photometric case). A future spectroscopic survey with enhanced sensitivity will clearly tighten the constraint. Dedicated deep observations in the main Athena X-IFU fields are foreseeable, also given the limited field of view of the instrument.
It is also worth to stress again that the prospects for background subtraction are much more favourable in the case of the spectroscopic sample than in the photometric case, and consequently the forecast can be considered more robust in the former case.

Before concluding, we would like to mention that the strategy here proposed for the Athena telescope could be undertaken, on a shorter timescale, by the X-ray Imaging and Spectroscopy Mission (XRISM)~\cite{2018SPIE10699E..22T}, expected to launch in 2022. XRISM has a reduced effective area with respect to Athena (by a factor of 45 at 1 keV, reducing to a factor of 6 at 7 keV and less at higher energies) but it can be able to improve the bounds derived in the eROSITA case for large sterile neutrino masses.

\section{Conclusions}
\label{sec:conclusions}

We have studied the X-ray signal produced by the $\nu_s\rightarrow\nu\, \gamma$ decay of sterile neutrino DM in cosmological structures.
For a given energy, such monochromatic signal is associated to a certain redshift slice.
Exploiting this property, we have investigated the cross correlation of the X-ray emission with catalogues of galaxies in the corresponding redshift range.

We have performed such line intensity mapping analysis for the eROSITA and Athena X-ray telescopes, and considering current and near future  photometric  and spectroscopic galaxy surveys.
Our main results are summarized in \Fig{fig:Sensitivities}.

eROSITA, which is currently in operation, will be able to slightly improve existing bounds and test the DM interpretation of the $3.55$ keV line excess.
Thanks to the improved effective area, Athena WFI will test a much larger and unexplored region of the parameter space.
Finally, the line intensity mapping technique can be fully exploited combining the superior spectral resolution of Athena X-IFU with spectroscopic surveys, like DESI.
The limited field of view of the X-IFU leads to sensitivities slightly worse than in the case of the WFI.
On the other hand, X-IFU spectroscopic observations can allow a more robust identification and characterization of the DM cosmological line.

Summarizing, we found that, in the near future, X-ray line intensity mapping can become a suitable technique to search for a sterile neutrino decay signal, complementary to observations of individual targets~\cite{Neronov:2015kca}.

\subsubsection*{Acknowledgments}

We thank E. Branchini for insightful comments.
MR acknowledges support by ``Deciphering the high-energy sky via cross correlation'' funded by the agreement ASI-INAF n. 2017-14-H.0, by the ``Departments of Excellence 2018 - 2022'' Grant awarded by MIUR (L. 232/2016), and by the research grant ``From  Darklight  to  Dark  Matter: understanding the galaxy/matter connection to measure the Universe'' No. 20179P3PKJ funded by MIUR.
MT acknowledge support from the INFN grant LINDARK and the research grant ‘The Dark Universe: A Synergic Multimessenger Approach’ No. 2017X7X85K funded by MIUR.
MR and MT acknowledge support from the project ``Theoretical Astroparticle Physics (TAsP)'' funded by the INFN. A.C. is supported by ``Generalitat Valenciana'' (Spain) through the ``plan GenT'' program (CIDEGENT/2018/019), by grants FPA2014-57816-P, PROMETEOII/2014/050 and SEV-2014-0398, as well as by the EU projects H2020-MSCA-RISE-2015 and H2020-MSCA-ITN- 2015//674896-ELUSIVES.

\bibliographystyle{hunsrt}
\bibliography{biblio}

\begin{thebibliography}{10}

\bibitem{Minkowski:1977sc}
Peter Minkowski.
\newblock {$\mu \to e\gamma$ at a Rate of One Out of $10^{9}$ Muon Decays?}
\newblock {\em Phys. Lett.}, 67B:421--428, 1977.

\bibitem{Mohapatra:1979ia}
Rabindra~N. Mohapatra and Goran Senjanovic.
\newblock {Neutrino Mass and Spontaneous Parity Nonconservation}.
\newblock {\em Phys. Rev. Lett.}, 44:912, 1980.
\newblock [,231(1979)].

\bibitem{Fukugita:1986hr}
M.~Fukugita and T.~Yanagida.
\newblock {Baryogenesis Without Grand Unification}.
\newblock {\em Phys. Lett.}, B174:45--47, 1986.

\bibitem{Brahmachari:1998kt}
Biswajoy Brahmachari and Rabindra~N. Mohapatra.
\newblock {Grand unification of the sterile neutrino}.
\newblock {\em Phys. Lett.}, B437:100--106, 1998, hep-ph/9805429.

\bibitem{Fritzsch:1974nn}
Harald Fritzsch and Peter Minkowski.
\newblock {Unified Interactions of Leptons and Hadrons}.
\newblock {\em Annals Phys.}, 93:193--266, 1975.

\bibitem{Nandi:1985uh}
S.~Nandi and U.~Sarkar.
\newblock {A Solution to the Neutrino Mass Problem in Superstring E6 Theory}.
\newblock {\em Phys. Rev. Lett.}, 56:564, 1986.

\bibitem{Mohapatra:1986bd}
R.~N. Mohapatra and J.~W.~F. Valle.
\newblock {Neutrino Mass and Baryon Number Nonconservation in Superstring
  Models}.
\newblock {\em Phys. Rev.}, D34:1642, 1986.
\newblock [,235(1986)].

\bibitem{Diaz:2019fwt}
A.~Diaz, C.~A. Argüelles, G.~H. Collin, J.~M. Conrad, and M.~H. Shaevitz.
\newblock {Where Are We With Light Sterile Neutrinos?}
\newblock 2019, 1906.00045.

\bibitem{Dentler:2018sju}
Mona Dentler, Álvaro Hernández-Cabezudo, Joachim Kopp, Pedro A.~N. Machado,
  Michele Maltoni, Ivan Martinez-Soler, and Thomas Schwetz.
\newblock {Updated Global Analysis of Neutrino Oscillations in the Presence of
  eV-Scale Sterile Neutrinos}.
\newblock {\em JHEP}, 08:010, 2018, 1803.10661.

\bibitem{Kopp:2011qd}
Joachim Kopp, Michele Maltoni, and Thomas Schwetz.
\newblock {Are There Sterile Neutrinos at the eV Scale?}
\newblock {\em Phys. Rev. Lett.}, 107:091801, 2011, 1103.4570.

\bibitem{Asaka:2005an}
Takehiko Asaka, Steve Blanchet, and Mikhail Shaposhnikov.
\newblock {The nuMSM, dark matter and neutrino masses}.
\newblock {\em Phys. Lett.}, B631:151--156, 2005, hep-ph/0503065.

\bibitem{Adhikari:2016bei}
M.~Drewes et~al.
\newblock {A White Paper on keV Sterile Neutrino Dark Matter}.
\newblock {\em JCAP}, 1701(01):025, 2017, 1602.04816.

\bibitem{Abazajian:2017tcc}
Kevork~N. Abazajian.
\newblock {Sterile neutrinos in cosmology}.
\newblock {\em Phys. Rept.}, 711-712:1--28, 2017, 1705.01837.

\bibitem{Boyarsky:2018tvu}
A.~Boyarsky, M.~Drewes, T.~Lasserre, S.~Mertens, and O.~Ruchayskiy.
\newblock {Sterile Neutrino Dark Matter}.
\newblock {\em Prog. Part. Nucl. Phys.}, 104:1--45, 2019, 1807.07938.

\bibitem{Dodelson:1993je}
Scott Dodelson and Lawrence~M. Widrow.
\newblock {Sterile-neutrinos as dark matter}.
\newblock {\em Phys. Rev. Lett.}, 72:17--20, 1994, hep-ph/9303287.

\bibitem{Shi:1998km}
Xiang-Dong Shi and George~M. Fuller.
\newblock {A New dark matter candidate: Nonthermal sterile neutrinos}.
\newblock {\em Phys. Rev. Lett.}, 82:2832--2835, 1999, astro-ph/9810076.

\bibitem{Kusenko:2006rh}
Alexander Kusenko.
\newblock {Sterile neutrinos, dark matter, and the pulsar velocities in models
  with a Higgs singlet}.
\newblock {\em Phys. Rev. Lett.}, 97:241301, 2006, hep-ph/0609081.

\bibitem{Petraki:2007gq}
Kalliopi Petraki and Alexander Kusenko.
\newblock {Dark-matter sterile neutrinos in models with a gauge singlet in the
  Higgs sector}.
\newblock {\em Phys. Rev.}, D77:065014, 2008, 0711.4646.

\bibitem{Khalil:2008kp}
Shaaban Khalil and Osamu Seto.
\newblock {Sterile neutrino dark matter in B - L extension of the standard
  model and galactic 511-keV line}.
\newblock {\em JCAP}, 0810:024, 2008, 0804.0336.

\bibitem{Merle:2013wta}
Alexander Merle, Viviana Niro, and Daniel Schmidt.
\newblock {New Production Mechanism for keV Sterile Neutrino Dark Matter by
  Decays of Frozen-In Scalars}.
\newblock {\em JCAP}, 1403:028, 2014, 1306.3996.

\bibitem{Shuve:2014doa}
Brian Shuve and Itay Yavin.
\newblock {Dark matter progenitor: Light vector boson decay into sterile
  neutrinos}.
\newblock {\em Phys. Rev.}, D89(11):113004, 2014, 1403.2727.

\bibitem{Abada:2014zra}
Asmaa Abada, Giorgio Arcadi, and Michele Lucente.
\newblock {Dark Matter in the minimal Inverse Seesaw mechanism}.
\newblock {\em JCAP}, 1410:001, 2014, 1406.6556.

\bibitem{Konig:2016dzg}
Johannes König, Alexander Merle, and Maximilian Totzauer.
\newblock {keV Sterile Neutrino Dark Matter from Singlet Scalar Decays: The
  Most General Case}.
\newblock {\em JCAP}, 1611(11):038, 2016, 1609.01289.

\bibitem{Merle:2015oja}
Alexander Merle and Maximilian Totzauer.
\newblock {keV Sterile Neutrino Dark Matter from Singlet Scalar Decays: Basic
  Concepts and Subtle Features}.
\newblock {\em JCAP}, 1506:011, 2015, 1502.01011.

\bibitem{Caputo:2018zky}
Andrea Caputo, Pilar Hernandez, and Nuria Rius.
\newblock {Leptogenesis from oscillations and dark matter}.
\newblock {\em Eur. Phys. J.}, C79(7):574, 2019, 1807.03309.

\bibitem{Kaneta:2016vkq}
Kunio Kaneta, Zhaofeng Kang, and Hye-Sung Lee.
\newblock {Right-handed neutrino dark matter under the $B - L$ gauge
  interaction}.
\newblock {\em JHEP}, 02:031, 2017, 1606.09317.

\bibitem{Asaka:2006ek}
Takehiko Asaka, Mikhail Shaposhnikov, and Alexander Kusenko.
\newblock {Opening a new window for warm dark matter}.
\newblock {\em Phys. Lett.}, B638:401--406, 2006, hep-ph/0602150.

\bibitem{Bezrukov:2009th}
F.~Bezrukov, H.~Hettmansperger, and M.~Lindner.
\newblock {keV sterile neutrino Dark Matter in gauge extensions of the Standard
  Model}.
\newblock {\em Phys. Rev.}, D81:085032, 2010, 0912.4415.

\bibitem{Nemevsek:2012cd}
Miha Nemevsek, Goran Senjanovic, and Yue Zhang.
\newblock {Warm Dark Matter in Low Scale Left-Right Theory}.
\newblock {\em JCAP}, 1207:006, 2012, 1205.0844.

\bibitem{Patwardhan:2015kga}
Amol~V. Patwardhan, George~M. Fuller, Chad~T. Kishimoto, and Alexander Kusenko.
\newblock {Diluted equilibrium sterile neutrino dark matter}.
\newblock {\em Phys. Rev.}, D92(10):103509, 2015, 1507.01977.

\bibitem{Gelmini:2019clw}
Graciela~B. Gelmini, Philip Lu, and Volodymyr Takhistov.
\newblock {Cosmological Dependence of Resonantly Produced Sterile Neutrinos}.
\newblock 2019, 1911.03398.

\bibitem{Gelmini:2019wfp}
Graciela~B. Gelmini, Philip Lu, and Volodymyr Takhistov.
\newblock {Cosmological Dependence of Non-resonantly Produced Sterile
  Neutrinos}.
\newblock 2019, 1909.13328.

\bibitem{Tremaine:1979we}
S.~Tremaine and J.~E. Gunn.
\newblock {Dynamical Role of Light Neutral Leptons in Cosmology}.
\newblock {\em Phys. Rev. Lett.}, 42:407--410, 1979.
\newblock [,66(1979)].

\bibitem{Lee:1977tib}
Benjamin~W. Lee and Robert~E. Shrock.
\newblock {Natural Suppression of Symmetry Violation in Gauge Theories: Muon -
  Lepton and Electron Lepton Number Nonconservation}.
\newblock {\em Phys. Rev.}, D16:1444, 1977.

\bibitem{Pal:1981rm}
Palash~B. Pal and Lincoln Wolfenstein.
\newblock {Radiative Decays of Massive Neutrinos}.
\newblock {\em Phys. Rev.}, D25:766, 1982.

\bibitem{Bulbul:2014sua}
Esra Bulbul, Maxim Markevitch, Adam Foster, Randall~K. Smith, Michael
  Loewenstein, and Scott~W. Randall.
\newblock {Detection of An Unidentified Emission Line in the Stacked X-ray
  spectrum of Galaxy Clusters}.
\newblock {\em Astrophys. J.}, 789:13, 2014, 1402.2301.

\bibitem{Boyarsky:2014jta}
Alexey Boyarsky, Oleg Ruchayskiy, Dmytro Iakubovskyi, and Jeroen Franse.
\newblock {Unidentified Line in X-Ray Spectra of the Andromeda Galaxy and
  Perseus Galaxy Cluster}.
\newblock {\em Phys. Rev. Lett.}, 113:251301, 2014, 1402.4119.

\bibitem{Boyarsky:2014ska}
Alexey Boyarsky, Jeroen Franse, Dmytro Iakubovskyi, and Oleg Ruchayskiy.
\newblock {Checking the Dark Matter Origin of a 3.53 keV Line with the Milky
  Way Center}.
\newblock {\em Phys. Rev. Lett.}, 115:161301, 2015, 1408.2503.

\bibitem{Riemer-Sorensen:2014yda}
Signe Riemer-Sørensen.
\newblock {Constraints on the presence of a 3.5 keV dark matter emission line
  from Chandra observations of the Galactic centre}.
\newblock {\em Astron. Astrophys.}, 590:A71, 2016, 1405.7943.

\bibitem{Dessert:2018qih}
Christopher Dessert, Nicholas~L. Rodd, and Benjamin~R. Safdi.
\newblock {Evidence against the decaying dark matter interpretation of the 3.5
  keV line from blank sky observations}.
\newblock 2018, 1812.06976.

\bibitem{Merloni:2012uf}
A.~Merloni et~al.
\newblock {eROSITA Science Book: Mapping the Structure of the Energetic
  Universe}.
\newblock 2012, 1209.3114.

\bibitem{Zandanel:2015xca}
Fabio Zandanel, Christoph Weniger, and Shin'ichiro Ando.
\newblock {The role of the eROSITA all-sky survey in searches for sterile
  neutrino dark matter}.
\newblock {\em JCAP}, 1509(09):060, 2015, 1505.07829.

\bibitem{Nandra:2013shg}
Kirpal Nandra et~al.
\newblock {The Hot and Energetic Universe: A White Paper presenting the science
  theme motivating the Athena+ mission}.
\newblock 2013, 1306.2307.

\bibitem{Creque-Sarbinowski:2018ebl}
Cyril Creque-Sarbinowski and Marc Kamionkowski.
\newblock {Searching for Decaying and Annihilating Dark Matter with Line
  Intensity Mapping}.
\newblock {\em Phys. Rev.}, D98(6):063524, 2018, 1806.11119.

\bibitem{Aghanim:2018eyx}
N.~Aghanim et~al.
\newblock {Planck 2018 results. VI. Cosmological parameters}.
\newblock 2018, 1807.06209.

\bibitem{Cappelluti:2017miu}
Nico Cappelluti et~al.
\newblock {The Chandra COSMOS legacy survey: Energy Spectrum of the Cosmic
  X-ray Background and constraints on undetected populations}.
\newblock {\em Astrophys. J.}, 837(1):19, 2017, 1702.01660.

\bibitem{Lieu:2015pit}
Maggie Lieu et~al.
\newblock {The XXL Survey IV. Mass-temperature relation of the bright cluster
  sample}.
\newblock {\em Astron. Astrophys.}, 592:A4, 2016, 1512.03857.

\bibitem{2014arXiv1409.4143B}
Esra {Bulbul}, Maxim {Markevitch}, Adam~R. {Foster}, Randall~K. {Smith},
  Michael {Loewenstein}, and Scott~W. {Randall}.
\newblock {Comment on ``Dark matter searches going bananas: the contribution of
  Potassium (and Chlorine) to the 3.5 keV line''}.
\newblock {\em arXiv e-prints}, page arXiv:1409.4143, Sep 2014, 1409.4143.

\bibitem{1953ApJ...117..134L}
D.~N. {Limber}.
\newblock {The Analysis of Counts of the Extragalactic Nebulae in Terms of a
  Fluctuating Density Field.}
\newblock {\em ApJ}, 117:134, January 1953.

\bibitem{Cooray:2002dia}
Asantha Cooray and Ravi~K. Sheth.
\newblock {Halo models of large scale structure}.
\newblock {\em Phys.Rept.}, 372:1--129, 2002, astro-ph/0206508.

\bibitem{Fornengo:2013rga}
Nicolao Fornengo and Marco Regis.
\newblock {Particle dark matter searches in the anisotropic sky}.
\newblock {\em Front. Physics}, 2:6, 2014, 1312.4835.

\bibitem{eRosita}
{eROSITA, https://www.mpe.mpg.de/455799/instrument}.

\bibitem{Merloni}
{A.Merloni,
  https://www.cosmos.esa.int/documents/332006/1402684/AMerloni\_t.pdf}.

\bibitem{Barret:2013bna}
D.~Barret et~al.
\newblock {Athena+: The first Deep Universe X-ray Observatory}.
\newblock 2013, 1310.3814.

\bibitem{xIFU}
{X-IFU, http://x-ifu.irap.omp.eu/the-x-ifu-in-a-nutshell/}.

\bibitem{Lotti:2014kda}
S.~Lotti, D.~Cea, C.~Macculi, T.~Mineo, L.~Natalucci, E.~Perinati, L.~Piro,
  M.~Federici, and B.~Martino.
\newblock {In-orbit background of X-ray microcalorimeters and its effects on
  observations}.
\newblock {\em Astron. Astrophys.}, 569:A54, 2014, 1410.3373.

\bibitem{Bilicki14}
M.~{Bilicki}, T.~H. {Jarrett}, J.~A. {Peacock}, M.~E. {Cluver}, and
  L.~{Steward}.
\newblock {Two Micron All Sky Survey Photometric Redshift Catalog: A
  Comprehensive Three-dimensional Census of the Whole Sky}.
\newblock {\em ApJ Suppl.}, 210:9, January 2014, 1311.5246.

\bibitem{SDSS2016}
R{\'o}bert {Beck}, L{\'a}szl{\'o} {Dobos}, Tam{\'a}s {Budav{\'a}ri},
  Alexander~S. {Szalay}, and Istv{\'a}n {Csabai}.
\newblock {Photometric redshifts for the SDSS Data Release 12}.
\newblock {\em MNRAS}, 460(2):1371--1381, Aug 2016, 1603.09708.

\bibitem{Abbott:2005bi}
T.~Abbott et~al.
\newblock {The dark energy survey}.
\newblock 2005, astro-ph/0510346.

\bibitem{Aghamousa:2016zmz}
Amir Aghamousa et~al.
\newblock {The DESI Experiment Part I: Science,Targeting, and Survey Design}.
\newblock 2016, 1611.00036.

\bibitem{Euclid2011}
R.~Laureijs et~al.
\newblock {Euclid Definition Study Report}.
\newblock {\em arXiv e-prints}, page arXiv:1110.3193, Oct 2011, 1110.3193.

\bibitem{Abell:2009aa}
Paul~A. Abell et~al.
\newblock {LSST Science Book, Version 2.0}.
\newblock page arXiv:0912.0201, 2009, 0912.0201.

\bibitem{Roach:2019ctw}
Brandon~M. Roach, Kenny C.~Y. Ng, Kerstin Perez, John~F. Beacom, Shunsaku
  Horiuchi, Roman Krivonos, and Daniel~R. Wik.
\newblock {NuSTAR Tests of Sterile-Neutrino Dark Matter: New Galactic Bulge
  Observations and Combined Impact}.
\newblock 2019, 1908.09037.

\bibitem{Boyarsky:2009ix}
Alexey Boyarsky, Oleg Ruchayskiy, and Mikhail Shaposhnikov.
\newblock {The Role of sterile neutrinos in cosmology and astrophysics}.
\newblock {\em Ann. Rev. Nucl. Part. Sci.}, 59:191--214, 2009, 0901.0011.

\bibitem{2018SPIE10699E..22T}
Makoto Tashiro et~al.
\newblock {Concept of the X-ray Astronomy Recovery Mission}.
\newblock In {\em Proceedings of the SPIE}, volume 10699 of {\em Society of
  Photo-Optical Instrumentation Engineers (SPIE) Conference Series}, page
  1069922, Jul 2018.

\bibitem{Neronov:2015kca}
A.~Neronov and D.~Malyshev.
\newblock {Toward a full test of the $\nu$MSM sterile neutrino dark matter
  model with Athena}.
\newblock {\em Phys. Rev.}, D93(6):063518, 2016, 1509.02758.

\bibitem{Ghirardini:2018byi}
V.~Ghirardini et~al.
\newblock {Universal thermodynamic properties of the intracluster medium over
  two decades in radius in the X-COP sample}.
\newblock {\em Astron. Astrophys.}, 621:A41, 2019, 1805.00042.

\bibitem{Eckert:2018mlz}
D.~Eckert et~al.
\newblock {Non-thermal pressure support in X-COP galaxy clusters}.
\newblock {\em Astron. Astrophys.}, 621:A40, 2019, 1805.00034.

\bibitem{Aird:2009sg}
J.~Aird et~al.
\newblock {The evolution of the hard X-ray luminosity function of AGN}.
\newblock {\em Mon. Not. Roy. Astron. Soc.}, 401:2531, 2010, 0910.1141.

\bibitem{Ptak:2007ae}
A.~Ptak, B.~Mobasher, A.~Hornschemeier, F.~Bauer, and C.~Norman.
\newblock {X-Ray Luminosity Functions of Normal Galaxies in the Great
  Observatories Origins Deep Survey}.
\newblock {\em Astrophys. J.}, 667:826--858, 2007, %.

\bibitem{Young:2012bm}
M.~Young et~al.
\newblock {Variability Selected Low-Luminosity Active Galactic Nuclei in the 4
  Ms Chandra Deep Field-South}.
\newblock {\em Astrophys. J.}, 748:124, 2012, 1201.4391.

\bibitem{Navarro:1996gj}
Julio~F. Navarro, Carlos~S. Frenk, and Simon~D.M. White.
\newblock {A Universal density profile from hierarchical clustering}.
\newblock {\em Astrophys.J.}, 490:493--508, 1997, astro-ph/9611107.

\bibitem{Prada:2011jf}
Francisco Prada, Anatoly~A. Klypin, Antonio~J. Cuesta, Juan~E. Betancort-Rijo,
  and Joel Primack.
\newblock {Halo concentrations in the standard LCDM cosmology}.
\newblock {\em Mon.Not.Roy.Astron.Soc.}, 428:3018--3030, 2012, 1104.5130.

\bibitem{Sheth:1999mn}
Ravi~K. Sheth and Giuseppe Tormen.
\newblock {Large scale bias and the peak background split}.
\newblock {\em Mon.Not.Roy.Astron.Soc.}, 308:119, 1999, astro-ph/9901122.

\bibitem{Takahashi:2012em}
Ryuichi Takahashi, Masanori Sato, Takahiro Nishimichi, Atsushi Taruya, and
  Masamune Oguri.
\newblock {Revising the Halofit Model for the Nonlinear Matter Power Spectrum}.
\newblock {\em Astrophys. J.}, 761:152, 2012, 1208.2701.

\bibitem{Zheng:2004id}
Zheng Zheng, Andreas~A. Berlind, David~H. Weinberg, Andrew~J. Benson,
  Carlton~M. Baugh, et~al.
\newblock {Theoretical models of the halo occupation distribution: Separating
  central and satellite galaxies}.
\newblock {\em Astrophys.J.}, 633:791--809, 2005, astro-ph/0408564.

\bibitem{Ammazzalorso:2018evf}
Simone Ammazzalorso, Nicolao Fornengo, Shunsaku Horiuchi, and Marco Regis.
\newblock {Characterizing the local gamma-ray Universe via angular
  cross-correlations}.
\newblock {\em Phys. Rev.}, D98(10):103007, 2018, 1808.09225.

\bibitem{Cuoco:2015rfa}
Alessandro Cuoco, Jun-Qing Xia, Marco Regis, Enzo Branchini, Nicolao Fornengo,
  and Matteo Viel.
\newblock {Dark Matter Searches in the Gamma-ray Extragalactic Background via
  Cross-correlations With Galaxy Catalogs}.
\newblock {\em Astrophys. J. Suppl.}, 221(2):29, 2015, 1506.01030.

\end{thebibliography}

\section*{Appendix A \\ X-ray emission from clusters}
\label{sec:cluster}

The X-ray cluster emission is due to bremsstrahlung radiation. The associated window function $W_{X_c}$ is given by (see also Ref.~\cite{Zandanel:2015xca})
\be
 W_{X_c}(E, z) = 
 \frac{ k_\mathrm{ff}\,\mathrm{exp} \left(- \frac{E\,(1+z)}{k_\mathrm{B}T_\mathrm{gas}(M)} \right)}{4\pi\,\sqrt{k_\mathrm{B}T_\mathrm{gas}(M)}\,E\,(1+z)} 
 \,\int d M \frac{d n}{d M}
  \int dV \rho_\mathrm{gas}^2(r|M) \;.
 \label{eq:Wcl}
\ee
To easy the physical interpretation, we can decompose it as $W_{X_c}=W_{X_c}^0\,\langle \delta_\mathrm{gas}^2\rangle$ with $W_{X_c}^0$ being defined as
\be
W_{X_c}^0(E, z) =
\frac{ k_\mathrm{ff}\,\left(\Omega_b\,\rho_c\right)^2}{4\pi\,\sqrt{k_{\rm B}T_{\rm gas}^0}\,E\,(1+z)} 
\exp{ \left(- \frac{E\,(1+z)}{k_{\rm B}T_{\rm gas}^0} \right)} \;,
 \label{eq:W0cl}
\ee
where $T_\mathrm{gas}^0$ is a reference temperature that can be arbitrarily chosen, and 
$k_\mathrm{ff} = \frac{32\pi e^6}{3 m_e c^2} \left(\frac{2\pi}{3k_\mathrm{B}m_e}\right)^{1/2}(4\pi\epsilon_0)^{-3}g_\mathrm{ff},$ where $g_\mathrm{ff}=1.1$ is the Gaunt factor for the
free-free emission.

The term $\langle \delta_\mathrm{gas}^2\rangle$ involves mass-dependent quantities:
\be
    \langle\delta_\mathrm{gas}^2\rangle = 
  \left( \frac{1}{\Omega_\mathrm{b}~\rho_\mathrm{c}} \right)^2 
  \int d M \frac{d n}{d M}\,f(T|M)\, 
  \int dV \rho_\mathrm{gas}^2(r|M) \;,
  \label{eq:d2cl}
\ee
where the function $f(T|M)$ is
\be
f(T|M)=\sqrt{\frac{T_\mathrm{gas}^0}{T_\mathrm{gas}(M)}}\,\mathrm{exp} \left(\frac{E\,(1+z)}{k_\mathrm{B}T_\mathrm{gas}^0}-\frac{E\,(1+z)}{k_\mathrm{B}T_\mathrm{gas}(M)}\right)\;.
\ee

$\rho_\mathrm{gas}$ and $T_\mathrm{gas}$ are the ICM density and temperature, respectively.
To describe them, we consider state-of-the-art models. More precisely, the shape of the ICM density $g(R/R_{500})$ is taken from Eq.~7 in Ref.~\cite{Ghirardini:2018byi} with values of the parameters from the best-fit in their Table 3.
The overall normalization is set through $M_{\mathrm{gas},500}=f_{\mathrm{gas},500}\,M_{500}=\int dV \rho_0\,g(R/R_{500})$, considering $f_{\mathrm{gas},500}=0.131$ \cite{Eckert:2018mlz}.
For the temperature-mass relation we adopt a power-law
\be
T_\mathrm{gas}(M_{500})=T_0\,\left(\frac{M_{500}\,E(z)}{M_0}\right)^\alpha\;,
\ee
setting the parameters of the relation from Ref.~\cite{Lieu:2015pit}: $\alpha=0.6$, $T_0=1$ keV and $M_0=10^{13.57}h_{70}^{-1} M_\odot$.

\begin{figure}[t]
\begin{center}
\includegraphics[width= 0.48 \textwidth]{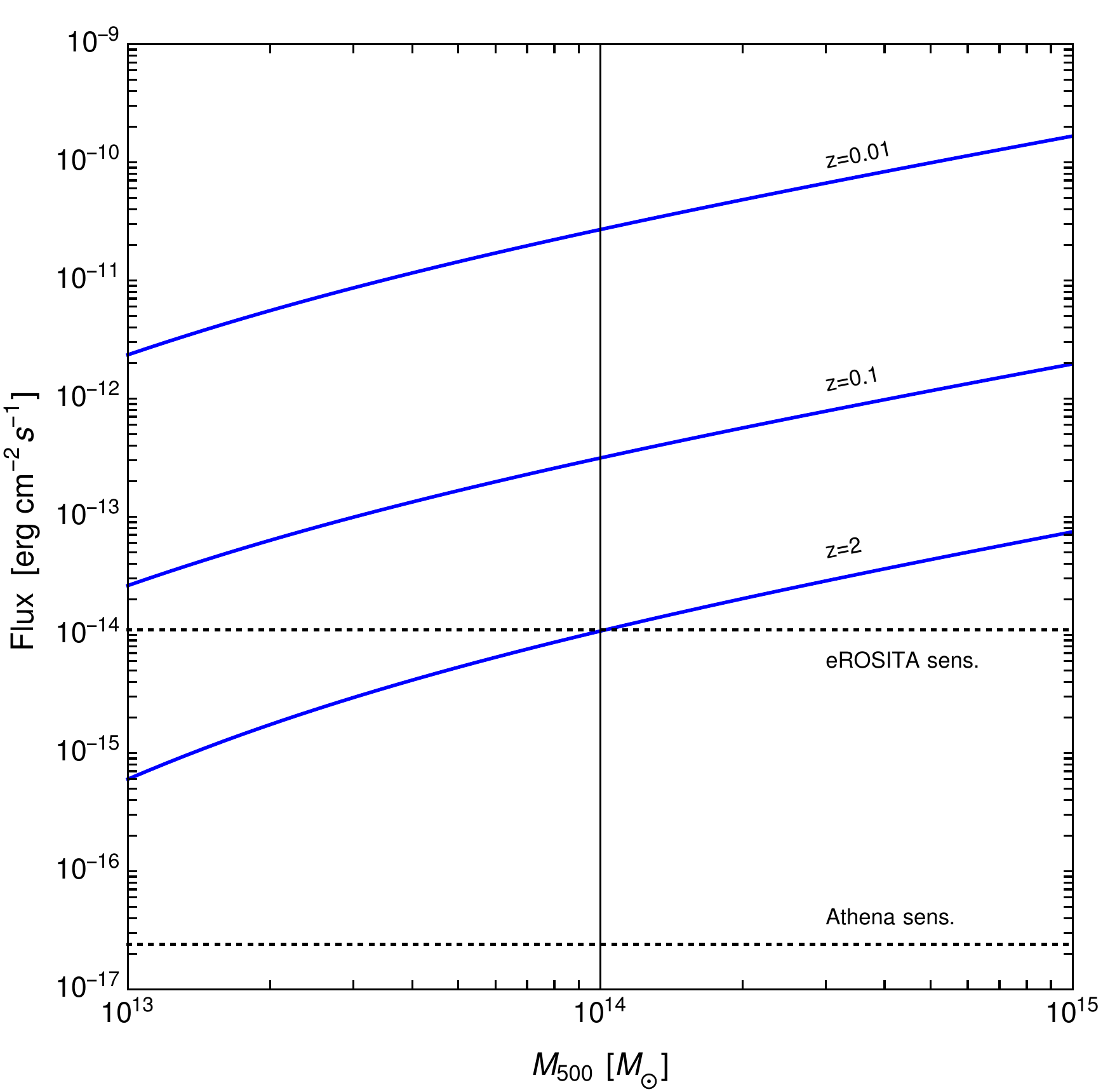}
\caption{X-ray flux from a cluster in the [0.5-2] keV band as a function of its mass and for three redshifts. We show also the source sensitivities of eROSITA and Athena telescopes.}
\label{fig:FluxClusters}
\end{center}
\end{figure}

In \Fig{fig:FluxClusters}, we show the predicted X-ray flux as a function of the cluster mass and for different redshift. All clusters with mass $M_{500}>10^{14}\, M_\odot$ are expected to be within the reach of eROSITA and Athena.
This means they can be masked and for this reason they are not included in our analysis of the correlation of the unresolved X-ray background.

\section*{Appendix B \\ X-ray luminosity function of AGN and galaxies}
\label{sec:Wastro}

The X-ray luminosity function of AGN is taken from the model of Ref.~\cite{Aird:2009sg}, calibrated on X-ray Chandra observations:
\be
\phi=K(z)\left[ \left(\frac{L}{L_*(z)}\right)^{\gamma_1}+ \left(\frac{L}{L_*(z)}\right)^{\gamma_2}\right] \;,
\label{eq:phiAGN}
\ee
with $\gamma_1=0.62,$ $\gamma_2=3.01$ and $K(z)=10^{-4.53-0.19\,(1+z)}.$
The characteristic luminosity is
\be
L_*(z)=L_0\left[ \left(\frac{1+z_c}{1+z}\right)^{p_1}+ \left(\frac{1+z_c}{1+z}\right)^{p_2}\right]^{-1} \;,
\label{eq:lumagn}
\ee
with $p_1=6.36,$ $p_2=-0.24,$ $z_c=0.75$ and $L_0=10^{44.77}$ erg s$^{-1}$.
The spectral energy distribution is taken to be a power-law with spectral index of 1.45, the best-fit value in Ref.~\cite{Cappelluti:2017miu}.

Concerning the galaxies, we adopt the model of Ref.~\cite{Ptak:2007ae}:
\be
\phi=\phi_* \left(\frac{L}{L_*}\right)^{1-\alpha} \exp \left[ -\frac{1}{2\sigma^2} \log_{10}^2\left(1+\frac{L}{L_*} \right)\right] \;,
\label{eq:phigal}
\ee

with $\alpha=1.43,$ $\sigma=0.72,$ $\phi_*=10^{-2.23}$ Mpc$^{-3}$ and $L_*=10^{39.74}\left(\frac{1+z}{1.25}\right)^{1.9}$ erg s$^{-1}$.
As for AGN, we consider a power-law spectral energy distribution, in this case with spectral index of 2~\cite{Young:2012bm}.

\section*{Appendix C \\ Three-dimensional power spectra}
\label{sec:ps}
In Eq.~\ref{eq:PShalo}, we described the general formalism used to estimate the 3D power spectra entering in our analysis. In order to apply them to each specific case, we need to specify the functions $\hat f$ (Fourier transforms of the field) and $b$ (bias of the source with respect to matter).

The case of auto correlation of X-rays from decaying DM (labeled with $\delta$) is given by
\bea
 P_{\delta,\delta}^{1h}(k,z) &=& \int_{M_{\rm min}}^{M_{\rm max}} dM\ \frac{dn}{dM} \tilde v_\delta(k|M)^2 \label{eq:PSdecdec}\\
 P_{\delta,\delta}^{2h}(k,z) &=& \left[\int_{M_{\rm min}}^{M_{\rm max}} dM\,\frac{dn}{dM} b_h(M)\,\tilde v_\delta(k|M) \right]^2\,P^{\rm lin}(k)\;,\nonumber
\eea
where $\tilde v_\delta(k|M)$ is the Fourier transform of $\rho_h(\bm x|M)/\bar \rho_{DM}$, with $\rho_h$ being the DM halo profile, for which we assume the NFW shape~\cite{Navarro:1996gj} with halo concentration from Ref.~\cite{Prada:2011jf}.
The bias of halos $b_h$ with respect to matter is taken from Ref.~\cite{Sheth:1999mn}, as well as the halo mass function $dn/dM$.
This estimate agrees very well with the non-linear matter power spectrum derived from N-body numerical simulations  \cite{Takahashi:2012em,Fornengo:2013rga}.
Let us remind that we are taking $M_{\rm max}=10^{14}\,M_\odot$, due to the masking of galaxy clusters.
For definiteness, we set $M_{\rm min}=10^{6}\,M_\odot$.

We consider AGNs and galaxies (labeled with $a$) as point-like sources, since their average size is smaller (or at most comparable) to the angular scales we analyze. Under this assumption, the 1-halo term is a Poissonian contribution (i.e., independent on $k$) which reads:
\bea
P_{a,a}^{1h}(z) &=& \int_{\mathcal{L}_{\rm min}(z)}^{\mathcal{L}_{\rm max}(z)} d\mathcal{L}\ \Phi_a(\mathcal{L},z)\,\left(\frac{\mathcal{L}}{\langle f_a \rangle}\right)^2
\eea
with $\langle f_a \rangle= \int d\mathcal{L}\,\Phi_a\,\mathcal{L}$.
This term is taken to be zero when cross correlating X-rays from AGNs with X-rays from galaxies.
The 2-halo term can be computed as $P_{a,a'}^{2h}(k,z)=\langle b_a(z)\rangle\,\langle b_{a'}(z)\rangle\, P_{\delta,\delta}^{2h}(k,z)$. We take the average bias $\langle b_a(z)\rangle$ from Ref.~\cite{Zandanel:2015xca}.

The power spectrum of cross correlation between X-rays from decaying DM and astrophysical sources is modeled through $P_{\delta,a}(k,z)=\langle b_a(z)\rangle P_{\delta,\delta}(k,z)$. This is exact for what concerns the 2-halo term, while just an approximation for the 1-halo component. We use this approximate description since the latter is just a highly subdominant contribution entering the covariance estimate.

In order to evaluate the correlation of galaxy catalogs with X-ray emitters, and since we adopt the halo model approach for the structure clustering, we need to describe how galaxies populate halos.\footnote{A different approach is to directly describe the galaxy clustering with an effective bias term, as done in Ref.~\cite{Zandanel:2015xca}.} To this aim, we employ the halo occupation distribution (HOD) formalism that provides the number of galaxies of a certain catalogue residing in a halo of mass $M$ at redshift $z$ and their spatial distribution. 
We follow the approach described, e.g., in Ref.~\cite{Zheng:2004id}, where the HOD is parameterized by distinguishing the contributions of central and satellite galaxies, $N_g=N_{\rm cen}+N_{\rm sat}$.
The details of the HOD model for the 2MPZ and SDSS catalogs are provided in Refs.~\cite{Ammazzalorso:2018evf} and \cite{Cuoco:2015rfa}.
For the on-going/future surveys DES, DESI and Euclid, we assume the same HOD as for SDSS. We expect this to be a good approximation, and small deviations from this reference model would have a negligible impact on our results.

For the catalog auto correlation, we refer the reader to e.g. Ref.~\cite{Cooray:2002dia}, while the power spectrum of cross correlation between decaying DM and a catalog of galaxies (labeled with $g$) can be written as:
\bea
 P_{g,\delta}^{1h}(k,z) &=& \int_{M_{\rm min}}^{M_{\rm max}^\delta} dM\ \frac{dn}{dM} \frac{\langle N_{g}\,\rangle}{\bar n_{g}}\tilde v_g(k|M) \tilde v_\delta(k|M) \label{eq:PSdecgal}\\
 P_{g,\delta}^{2h}(k,z) &=& \left[\int_{M_{\rm min}}^{M_{\rm max}^g} dM\,\frac{dn}{dM} b_h(M) \frac{\langle N_{g}\rangle}{\bar n_{g}} \tilde v_g(k|M) \right] \nonumber \\
&\times& \left[\int_{M_{\rm min}}^{M_{\rm max}^\delta} dM\,\frac{dn}{dM} b_h(M)\,\tilde v_\delta(k|M) \right]\,P^{\rm lin}(k)\;,\nonumber
\eea
where the notation emphasises that the maximum mass considered for the X-ray signal is different (smaller) than for galaxy catalogs.

The product $\langle N_{g}\rangle\,\tilde v_g(k|m)$ is the Fourier transform of $\langle N_{\rm cen}(M)\rangle\,\delta^3(\bm x )+\langle N_{\rm sat}(M)\rangle\,\rho_h(\bm x|M)/M$. 
%One can see that $\langle N_{g}\rangle\,\tilde v_g(k=0|m)=\langle N_{g}\rangle$.
The average number of galaxies at a given redshift is given by $\bar n_{g}(z)=\int dM\,dn/dM\, \langle N_{g}\rangle$.

The power spectrum of cross correlation between X-rays from astrophysical sources (AGN and galaxies) and galaxy catalogs is approximated with $P_{g,a}(k,z)=\langle b_a(z)\rangle P_{g,\delta}(k,z)$, following the same reasoning discussed above for the case of $P_{\delta,a}$.

\end{document}